\documentclass[11pt,authoryear,leqno,preprint]{elsarticle}
\usepackage{mathrsfs}
\usepackage{amsfonts}
\usepackage{amsmath}
\usepackage{amssymb}
\usepackage{amsthm}
\usepackage{amstext}
\usepackage{amsopn}
\usepackage{float}
\usepackage{color}
\usepackage{ulem}
\usepackage{multirow}
\usepackage{hyperref}
\usepackage{graphicx}

\DeclareGraphicsExtensions{.eps, .pdf, .jpg, .jpeg, .tif}

\newtheorem{theorem}{Theorem}[section]

\theoremstyle{definition}

\theoremstyle{remark}

\newcommand{\ds}{\displaystyle}
\newcommand{\la}{\lambda}
\newcommand{\noi}{\noindent}

\newcommand{\R}{{\mathbb R}}

\begin{document}

\begin{frontmatter}

\title{ Bistability in a Differential Equation Model of  \\Oyster Reef Height and Sediment
Accumulation}
\author[math]{William C.~Jordan-Cooley}
\author[vims]{Romuald N.~Lipcius}
\author[apsc]{Leah B.~Shaw}
\author[vims]{Jian Shen}
\author[math]{Junping Shi\footnote{Corresponding Author. Telephone: 1-757-221-2030, Fax: 1-757-221-7400, Email: \texttt{jxshix@wm.edu}}}

\address[math]{Department of Mathematics, College of William \& Mary, Williamsburg, Virginia, 23187-8795, USA}
\address[vims]{Virginia Institute Marine Science, College of William \& Mary,
 Gloucester Point, Virginia, 23062, USA}
\address[apsc]{Department of Applied Science, College of William \& Mary, Williamsburg, Virginia, 23187-8795, USA}


\begin{abstract}
Native oyster populations in Chesapeake Bay have been the focus of three decades of restoration attempts, which have generally failed to rebuild the populations and oyster reef structure. Recent restoration successes and field experiments suggest that high-relief reefs offset heavy sedimentation and promote oyster survival, disease resistance and growth, in contrast to low-relief reefs which degrade in just a few years. These findings suggest the existence of alternative stable states in oyster reef populations. We developed a mathematical model consisting of three differential equations that represent volumes of live oysters, dead oyster shells (= accreting reef), and sediment. Bifurcation analysis and numerical simulations demonstrated that multiple nonnegative equilibria can exist for live oyster, accreting reef and sediment volume at an ecologically reasonable range of parameter values; the initial height of oyster reefs determined which equilibrium was reached. This investigation thus provides a conceptual framework for alternative stable states in native oyster populations, and can be used as a tool to improve the likelihood of success in restoration efforts.
\vspace{0.1in}

\end{abstract}

\end{frontmatter}

\section{Introduction}

\subsection{Decline and restoration of native oyster populations}

In the past century, the native Chesapeake Bay oyster, \textit{Crassostrea virginica}, a dominant ecosystem engineer \citep{Cerco07}, has dropped to approximately 1\% of its previous abundance due to overfishing and habitat degradation \citep{Rothschild94}. Harvests peaked in 1884 at 615,000 metric tons, but in 1992 the harvest was only 12,000 metric tons. Additionally, human activities on land increased the flow of sediment into the bay's waters, which weakened physiological health, lowered fecundity and raised mortality of oysters \citep{Newell88, Rothschild94, Lenihan99}. Exacerbating the situation, the physical profile of reefs has been leveled by fishers exploiting oyster reefs \citep{Rothschild94}, which places the oysters lower in the water column where water flow is reduced and sediment accumulation rates are highest, thereby suffocating oysters \citep{Newell88, Lenihan99}.

Efforts to restore native oyster populations have been extensive but largely ineffectual \citep{Ruesink05}. However, a recent restoration effort in the Great Wicomico River has yielded promising results. In $2004$, the Army Corps of Engineers created nearly $40$ ha of reef in the Great Wicomico River (Chesapeake Bay) consisting of oyster shell planted at different reef heights \citep{Schulte09}. The field experiment featured high-relief reefs built at an average of $25-42$ cm in height, low-relief reefs at $8-12$ cm, and controls of unrestored bottom. After three years, in $2007$, the higher reefs were considerably more successful. Mean oyster density was fourfold higher on high-relief reefs, about $1000$ oysters per $m^2$, than on low-relief reefs. Moreover, the high-relief reefs were robust in architecture and resistant to natural disturbances, whereas the low-relief reefs were heavily sedimented and apparently on a trajectory to a degraded state. These differing outcomes suggested the potential for alternative stable states driven by the initial condition of reef height.

\subsection{Alternative stable states}

The dramatic decline in the oyster population as well as the marked difference in success of the high-relief and low-relief reefs may be explained in the context of catastrophic shifts and bistability (\textit{i.e.}, alternative stable states). Some ecosystems exhibit precipitous shifts in state without correspondingly dramatic changes in external conditions \citep{Scheffer01}. Gradual changes in the external conditions of a system produce correspondingly gradual alterations in state variables until sudden drastic changes occur in the state variable. After a degraded state has been reached, return to the external (\textit{i.e.}, environmental) conditions present before the change does not induce reversion of the system to the pre-shift state. These systems exhibit bistability, which is one manifestation of the existence of multiple equilibria for a range of constant environmental conditions \citep{Scheffer01, Dong02, Guill09}.

Alternative (= multiple) stable states in communities have been triggered by environmental disturbances in various ecosystems including coral reefs \citep{Hughes94}, lakes \citep{Carpenter99}, grasslands \citep{Koppel97}, and kelp forests \citep{Konar03}. In mollusks (\textit{e.g.}, mussels, clams, and oysters) alternative stable states have been documented in beds of the horse mussel  \textit{Atrina zelandica} in New Zealand \citep{Coco06} and the blue mussel along the northeast Atlantic coast of North America \citep{Petraitis09}. Despite the likelihood of multiple stable states in marine species, there have been few mathematical models of this process, particularly for mollusks \citep{Petraitis10}. Moreover, for native oyster populations, the mathematical models of population dynamics have usually emphasized linear interactions without the potential for alternative stable states \citep{Powell06}, whereas biologically realistic models of oyster filtration have integrated nonlinear biological processes \citep{Cerco07}. Consequently, our mathematical formulation represents an advance in the mathematical modeling of alternative stable states in the population dynamics of exploited marine species such as the oyster, and thereby advances the theoretical underpinnings for ecological restoration of marine species.

\subsection{Feedback mechanisms for alternative stable states}

Alternative stable states are generally due to one or more feedback mechanisms \citep{Scheffer01, Guill09}. In the Chesapeake Bay, oysters filter phytoplankton (microscopic algae) and sediment flowing onto reefs, which lowers turbidity levels and the incidence of low dissolved oxygen conditions \citep{Newell88}. Historically, massive reductions in oyster biomass and degradation of the reef matrix contributed to increasing sediment in the water column and low dissolved oxygen on the bottom of the bay. Additionally, oysters encounter a greater proportion of the sediment in the water column when they are closer to the bottom \citep{vanrijn}, which occurs with reductions in the vertical relief of reefs. The sediment negatively affects oysters by causing them to expend energy to filter it, thereby increasing susceptibility to disease and mortality rates, while decreasing growth and reproduction. Raising the oysters in the water column can lead to higher fecundity and decreased mortality from reductions in turbidity and elevated filtration rates \citep{Lenihan99, Ruesink05}.

A well known ecosystem in which feedback mechanisms lead to alternative stable states is that of shallow lakes, which have ecosystem properties similar to those of Chesapeake Bay. In shallow lakes, aquatic vegetation dampens resuspension of sediment, reduces nutrients in the water column, and provides protection from fish predation for zooplankton that feed on phytoplankton \citep{Scheffer09}. Fish control zooplankton and resuspend sediment and nutrients by disturbing the lake floor.  However, aquatic vegetation and zooplankton have been depleted by herbicides and pesticides. The reduction in vegetation leads to an increase in nutrients in the water column which increases phytoplankton, which in turn feed on the nutrients. Declines of zooplankton result in unchecked growth of phytoplankton. A dramatic increase in phytoplankton precludes light from reaching the lake floor which causes the vegetation to decline further. This exposes zooplankton to increased predation, which in turn leads to lower predation pressure on phytoplankton \citep{Scheffer09}. To return the lakes to a state of high vegetation and controlled phytoplankton abundance, zooplankton populations must be rebuilt to a critical level such that they can control phytoplankton, and thus facilitate the profusion of vegetation that provides protection from fish predation. By temporarily reducing fish abundance, zooplankton can proliferate and feed on the abundant phytoplankton. Once phytoplankton are reduced, the vegetation can become re-established. The increased zooplankton eventually allows the fish population to rebound, restoring the system to the pristine stable state \citep{Scheffer09}.

We hypothesize that the oyster reef system is somewhat analogous to that of shallow lakes. Oysters can control the volume of sediment while consuming phytoplankton \citep{Newell88}, but they must first be provided with optimal reef features, particularly an elevated reef height. This enhances recruitment of young oysters \citep{Schulte09}, which subsequently grow to a dense spawning stock that further filters phytoplankton and sediments from the water column and preclude the accumulation of sediments on the reef. Without the initial reef height it has been postulated that the low-relief reef structure inhibits high recruitment, resulting in sparsely distributed oysters that cannot keep pace with the sediments and phytoplankton, eventually leading to a heavily silted, degrading reef. We now provide the theoretical underpinnings for this interaction between oysters, reef height and sediment, and demonstrate that alternative stable states are feasible for oyster reef populations.

\subsection{Objectives}

We construct an ordinary  differential equation model to demonstrate that the mechanisms involved in the interaction of oysters, reef height and sediment produce bistability, which provides an explanation for  the success of high-relief reefs and failure of low-relief reefs\citep{Schulte09} in the context of multiple stable states. Bifurcation theory is used to identify parameter ranges that produce bistability in the model. The bistable structure is sensitive to the initial values of the system. A small perturbation of the initial value can change the eventual outcome from one stable state to a different one, which is termed a \lq\lq hair-trigger effect\rq\rq. In general the basins of attraction of the two stable states are only separated by a surface in the phase space \citep{Jiang09}. We demonstrate our model's sensitivity to initial conditions via numerical simulations. The mathematical model of oyster and sediment is presented and motivated in Section \ref{sec3}. Analytic bifurcation results are given in Section \ref{sec4a}, and numerical simulations for specific parameter values are in Section \ref{sec4b}. In Section \ref{sec6} we discuss the implications for alternative stable states in oysters for the likelihood of success of native oyster restoration, with emphasis on the Chesapeake Bay ecosystem.


\section{Methods: mathematical model}\label{sec3}

\subsection{Basic elements of the model}

We  model the rate of change of live oysters, dead oyster shells, and deposited sediment volume with respect to time, $t$, measured in years. Our state variables are measured in volume per $m^2$ of sea floor, so they can be converted to heights measured from the sea floor by dividing by a unit area. The feedback interaction between live oysters and sediment occurs as follows. Live oysters follow logistic growth but are negatively affected by sediment volume. We  introduce a function to represent the proportion of oysters above the level of suffocating sediment. The change in dead oyster shell volume is due to the death of live oysters minus a degradation rate proportional to dead oyster volume. The volume of sediment deposited on the reef depends on the position of the reef in the water column and on filtration by live oysters. We next give a detailed derivation for each differential equation. Variables are summarized in Table \ref{tab1} and parameters in Table \ref{tab2}.

\begin{table}[h]
\centering
\begin{tabular}{|c|c|c|}
  \hline
  \textit{Variable} & \textit{Description} & \textit{Units} \\
  \hline
  $t$ & time & years \\
  \hline
  $O(t)$ & live oysters & $m^3$ \\
  \hline
  $B(t)$ & dead oyster shells & $m^3$ \\
  \hline
  $S(t)$ & sediment on reef & $m^3$  \\
  \hline
\end{tabular}
\caption{Model variables.}\label{tab1}
\end{table}

\subsection{Proportion of oysters unaffected by sediment}

We  introduce a function $f$ that represents the proportion of oysters not affected by sediment.  The input, $d_0$, represents the volume of the live oysters ($O$) and dead oyster shells ($B$) not affected by sediment ($S$).  Hence we define
\begin{equation}\label{4}
    d_0 = O + B - S.
\end{equation}
Note that the volume is essentially height multiplied by a unit area, so $d_0$ represents the height of the oysters above the sediment. In this we make a simplifying assumption that there is a 1:1 relation between oyster or shell volume and sediment volume. Future work will model in more detail how sediment volume is affected by reef geometry.

The proportion of oysters not affected by sediment is  an increasing function of $d_0\in (-\infty,\infty)$ with a sigmoid shape bounded by $0$ and $1$. We denote this function by $f_1(d_0)$. When $d_0 = O$, the dead oyster shells are covered by sediment but the live oysters are not, \textit{i.e.}, when $B = S$. In this situation, live oysters are not affected by sediment and $f_1(O)\approx 1$. When $d_0 = 0$, both live and dead oysters are covered by sediment, and $f_1(0) \approx 0$.  As $d_0$ approaches the positive and negative limits, the approximations are equalities. We can assume that  $f_1(d_0)$ is a sigmoid function with the only reflection point at the midpoint of the interval $[0,O]$ and $f_1(O/2)=1/2$. By the replacement $d=d_0-O/2$, the function $f_1(d_0)$ becomes a function of:
\begin{equation}\label{4a}
    d=d_0-\frac{O}{2}=\frac{O}{2}+B-S.
\end{equation}
Then we define $f(d)=f_1(d_0-O/2)$. In summary, $f=f(d)$ satisfies
\begin{equation}\label{5}
    f'(d)>0, \;\;\; f(0)=\frac{1}{2}, \;\;\;  \lim_{d\to-\infty}f(d)=0, \;\;\; \text{ and } \lim_{d\to\infty}f(d)=1.
 \end{equation}
The function $f$ was devised to make $f(d)\approx 0$ when $d<-k_1/2$ and $f(d)\approx 1$ when $d>k_1/2$, where $k_1$ is a parameter proportional to the carrying capacity of the live oysters. Thus, $O/2$ oysters must be covered by sediment before their performance is affected by sediment; \textit{i.e.}, we assume that more than half of any individual oyster must be covered by sediment for its performance to suffer.

\subsection{Live oyster volume}

The change in live oyster volume $O(t)$ is represented by the differential equation
\begin{equation}
\label{1a} \frac{d O}{d t} = r O f(d) \left(1 -
\frac{O}{K}\right)-\mu f(d) O - \epsilon(1 - f(d)) O.
\end{equation}
Here the live oysters are assumed to follow logistic growth, $K$ is the carrying capacity, and $r$ represents the intrinsic rate of increase. The oyster population increases at a negative density-dependent rate until the population reaches $K$.  In the second term of the equation, $\mu$ represents mortality due to predation and disease. Both terms are scaled by $f(d)$ because oysters covered in sediment do not reproduce, and are assumed to die due to suffocation by sediment rather than by predation or disease. The third term represents the decrease in live oyster volume as a result of sediment; $\epsilon$ is the death rate of oysters covered by sediment. As $f(d)$ goes to $1$, this term goes to zero. However, as $f(d)$ becomes smaller, the term begins to exert a greater effect.

\subsection{Dead oyster (shell) volume}

The second differential equation represents the change of dead oyster shell volume $B(t)$ as
\begin{equation}
\label{2a} \frac{d B}{d t} = \mu f(d) O + \epsilon(1 - f(d))
O-\gamma B.
\end{equation}
The first two terms are directly from the death of live oysters in Equation \eqref{1a}, and the third term is the loss of dead oyster volume due to degradation of shell. This loss is proportional to the volume of dead oysters at the rate $\gamma$. Note that the term $rO^2f(d)/K$ in \eqref{1a} is not included in \eqref{2a} as it is a reduction in growth rate rather than a loss term, and it does not increase the dead oyster volume as do the other two terms.


\subsection{Sediment accumulation}

The system of differential equations is completed by a third equation describing the change of sediment volume $S(t)$ as:
\begin{equation}
\label{3a} \frac{dS}{dt} = - \beta S+Cge^{-\frac{FO}{Cg}} .
\end{equation}
Here the first term is the rate of sediment erosion, which is proportional to the volume of deposited sediment with a rate $\beta$; the second term is the rate of sediment deposition. The sediment deposition rate in the absence of oysters is $Cg$, where $C$ is a maximum possible deposition rate and $g$ is a modification that depends on reef height $O+B$. The deposition rate is at its maximum at the sea floor, and decreases as the reef height in the water column increases \citep{Nielsen92}. Hence with the reef height represented by $x=O+B$, we assume that the function $g(x)=g(O+B)$ satisfies
\begin{equation}\label{g}
g(0)=1, \;\; g'(x)\le 0 \,\; \text{ for } \;\; x\ge 0, \;\;\;\; \text{ and } \;\;\;\;
\lim_{x\to\infty}g(x)=0.
\end{equation}
In this model formulation we assume that biodeposition is a constant, minor fraction of sediment deposition \citep{Cerco07}, and which would not alter the qualitative results of the modeling. Other processes that are not simulated explicitly, such as increased deposition due to feces and pseudofeces at the bottom, are parameterized by the erosion and burial rates. In future, more complex formulations we will integrate biodeposition into the model to generate results that are more accurate quantitatively.

In the presence of live oysters, the deposition term should be reduced by a multiplicative factor due to filtration. The filtration rate $F$ per unit oyster volume  depends on the height-dependent sediment concentration, which is proportional to $Cg$. The rate $F$ scales
linearly with $Cg$ when $Cg$ is small, reaches a peak $F_0$ at some optimal sediment concentration, and beyond this threshold, it
decreases as oyster gills become increasingly clogged \citep{Jordan87}. Hence $F=F(y)=F(Cg)$ satisfies
\begin{equation}\label{F}
\begin{split}
&F(0)=0; \;\; \lim_{y\to\infty}F(y)=0; \;\; \mbox{and there
exists} \;\; y_0>0 \;\; \mbox{such that}\\
&F'(y)>0 \;\; \mbox{for} \;\; 0<y<y_0, \;\; F'(y)<0 \;\; \mbox{for} \;\;  y>y_0, \;\; \mbox{and}
\;\; F(y_0)=F_0.
\end{split}
\end{equation}
Both $g(x)$ and $F(y)$ are positive and continuously differentiable functions on $[0,\infty)$.

The derivation of \eqref{3a} begins with a mass balance of sediment deposition in a control volume at the bottom \citep{Chapra97, Ji08}:
\begin{equation}\label{5a}
    \frac{d(V C_s)}{dt}=\Omega_s A  C_b -V_e A  C_s- V_b A  C_s-V_f  O,
\end{equation}
where $\Omega_s$ is settling velocity $(m/day)$, $V_e$ is erosion velocity $(m/day)$, $V_b$ is burial velocity $(m/day)$ of sediment that is consolidated in the bottom and lost from the system, $C_s$ is sediment concentration at the bottom $(g/m^3)$, $C_b$ is the sediment concentration above the bottom sediment layer $(g/m^3)$, $V$ is the control volume $(m^3)$, $A$ is the surface area $(m^2)$ of the control volume, $V_f$ is the filtration rate by oysters $(g/(m^3 \cdot day))$, and $O$ is live oyster volume $(m^3)$.

We introduce a mean erosion rate $v_e$ $(1/day)$ and burial rate $v_b$ $(1/day)$, which can be considered to be a re-scaling of the erosion and burial velocities $V_e$ and $V_b$ by the mean depth as follows:
\begin{equation}\label{5b}
    \frac{d(V C_s)}{dt}=\Omega_s A C_b -v_e V C_s- v_b V C_s-V_f O.
\end{equation}

The bottom sediment concentration can be expressed as sediment density $\rho$ and the porosity $\phi$ as $C_s = \rho(1-\phi)$ \citep{Chapra97}. We assume that the mean porosity, $\bar{\phi}$, is a constant and divide $\eqref{5b}$ by $\rho(1-\bar{\phi})$. Let $S =
\ds\frac{V(1-\phi)}{1-\bar{\phi}}$ be the sediment volume with porosity normalized by the mean porosity to quantify the volume of
unconsolidated sediment deposition at the bottom, and let $\omega_s = \ds\frac{\Omega_s A}{\rho(1-\bar{\phi})}$ and $v_f = \ds\frac{V_f}{\rho(1-\bar{\phi})}$. Then $\eqref{5b}$ can be written as
\begin{equation}\label{5c}
    \frac{dS}{dt}=\omega_s C_b -v_e S- v_b S-v_f O.
\end{equation}

Rearrangement of the terms in Equation \eqref{5c} yields
\begin{equation}
    \frac{dS}{dt}=\omega_s C_b \left(1-\frac{v_f O}{\omega_s C_b}\right) - (v_e + v_b) S.
\end{equation}
Here the first term represents the sediment deposition modified by oyster filtration, and the second term is the combined loss of sediment
due to erosion or burial. Because the modified deposition term could be negative for large values of $O$, we instead replace the decreasing linear function $\ds 1-\frac{v_fO}{\omega_S C_b}$ by a decreasing nonlinear function $\ds \exp\left(\frac{-v_fO}{\omega_S C_b}\right)$ whose linearization is $\ds 1-\frac{v_fO}{\omega_S C_b}$. Now we rename
\begin{equation*}
\omega_S C_b=C g, \;\; \text{ and } \;\; v_f=F,
\end{equation*}
where $C$ is a constant representing the maximum deposition rate, and $g$ is a decreasing function of $O+B$ with maximum $g(0)=1$; then we obtain \eqref{3a}. The estimates of $C$ and $g$ come from data of $\omega_S$ and $C_b$; information on $v_f$ can be used to determine the form and parameter values of $F$; and, $v_e$ and $v_b$ determine $\beta$.

Equation \eqref{3a} has properties reflecting the desired qualities of the system. When $O \rightarrow 0$, $\dot{S} \sim Cg - \beta S$ (deposition and erosion without oysters). When $O$ is small, from a Taylor expansion of the exponential we have $\dot{S} \sim Cg -FO - \beta S$ so the deposition rate is reduced linearly by the amount $FO$ that oysters filter out of the system. When $O \rightarrow \infty$, $\dot{S} \sim -\beta S$ (no deposition, only erosion). Finally, when the filtration rate per unit oyster volume $F \rightarrow 0$, either because there is too little sediment to filter or the oysters are being choked by it, $\dot{S} \sim Cg-\beta S$.

\subsection{Full model}

In summary, we propose the following set of differential equations to model oyster population, oyster reef, and sediment volumes:
\begin{align}
    \label{1}
\frac{d O}{d t} &= r O f(d) \left(1 - \frac{O}{k}\right)-\mu f(d) O - \epsilon(1 - f(d)) O,\\
\label{2}
\frac{d B}{d t} &= \mu f(d) O  + \epsilon(1 - f(d)) O- \gamma B,\\
\label{3} \frac{dS}{dt} &= - \beta S+Cge^{-\frac{FO}{Cg}},
\end{align}
where the quantities $d$, $f(d)$, $g=g(O+B)$ and $F=F(Cg)$ satisfy \eqref{5}, \eqref{g} and \eqref{F}, respectively. A set of functions satisfying these conditions will be given in Section 4. The parameters of the system \eqref{1}-\eqref{3} are summarized in Table $\ref{tab2}$. The last two parameters $h$ and $\eta$ are specific to the choice of $f$ and $g$, which will be explained in Section \ref{sec4}.

{\small
\begin{table}[htb]
\centering
\begin{tabular}{|c|c|c|c|c|}
  \hline
  Parameter  & Meaning & Units &  Value & Reference\\
  \hline
  $r$ & instantaneous rate of increase & {$year^{-1}$} & {$0.7-1.3$} & 1\\
   \hline
  $K$ & oyster carrying capacity &  $m^3$  & $0.1-0.3$ & 2\\
   \hline
  $\mu$ & mortality rate due to  &  $year^{-1}$ & $ 0.2-0.6 $ & 3\\
  & predation and disease  & & & \\
   \hline
  $\epsilon$ & mortality rate due to sediment & $year^{-1}$ & $0.94$ & 4\\
     \hline
   $\gamma$ & oyster shell degradation rate & $year^{-1}$ & $0.5-0.9 $ & 4\\
   \hline
  $F_0$ & maximum sediment filtration rate & $year^{-1}$ & $ 1 $ & 5\\
   \hline
  $C$ & maximum sediment deposition rate & $m^{3}\cdot year^{-1}$ & $ 0.04-0.08 $ & 6, 7\\
   \hline
   \multirow{2}{*}
  {$y_0$} &  sediment amount where & {$ year\cdot m^{-3} $} & {$0.02$} & 5\\
  &  the filtration is maximum & & & \\
   \hline
  $\beta$ & sediment erosion rate & $m^{-3}$ &  $0.02-0.04$ &  6, 7\\
   \hline
   $h$ & scaling factor & $m^{-3}$ & $10-30$ & \\
   \hline
   \multirow{2}{*}
{$\eta$} & decay rate of sediment deposition  & {$m^{-3}$} & {$3.33$} & 8\\
& on the reef height & & &\\
   \hline
\end{tabular}
\caption{Table of parameters in the equations. References: 1 \citep{EIS09}, 2 \citep{Schulte09}, 3 \citep{Pow09}, 4 \citep{Smith05}, 5 \citep{Jordan87}, 6 \citep{Kniskern03}, 7 \citep{Chapra97}, 8 \citep{vanrijn}.}\label{tab2}
\end{table}
}

\section{Results}\label{sec4}

\subsection{Bifurcation analysis}
\label{sec4a}

We are interested in determining the values of parameters and state variables at which the change in the state variables is equal to zero (\textit{i.e.} a steady state reef-sediment system). We consider the equilibrium solutions of our model \eqref{1}-\eqref{3}, which satisfy
\begin{align}
\label{7}
0 &= r O f(d) \left(1 - \frac{O}{K}\right)-\mu f(d) O - \epsilon(1 - f(d)) O, \\
\label{8}
0 &= \mu f(d) O  + \epsilon(1 - f(d)) O- \gamma B,\\
\label{9} 0 &= - \beta S+Cge^{-\frac{FO}{Cg}} .
\end{align}

A trivial solution of $\eqref{7}-\eqref{9}$ where $O = B = 0$ and $S =C/\beta$ is an equilibrium solution representing the extinction of the oyster population and the accumulation of sediment limited only by erosion. We will now solve the system in search of nontrivial solutions where $O > 0$, $B > 0$ and $S>0$.

We describe a procedure of reducing the equations \eqref{7}-\eqref{9} to a single equation. From \eqref{7}, we obtain (assuming $O>0$)
\begin{equation}\label{10}
    f(d)=\frac{\epsilon K}{K(r-\mu+\epsilon)-rO};
\end{equation}
and similarly from \eqref{8}, we obtain
\begin{equation}\label{11}
    f(d)=\frac{\gamma B-\epsilon O}{(\mu-\epsilon)O}.
\end{equation}
From \eqref{10} and \eqref{11}, we can have an equation of $O$ and $B$ only:
\begin{equation}\label{13}
\frac{\epsilon K}{K(r-\mu+\epsilon)-rO}=\frac{\gamma B-\epsilon O}{(\mu-\epsilon)O},
\end{equation}
and $B$ can be solved from \eqref{13} as
\begin{equation}\label{14}
    B=\frac{r\epsilon O(K-O)}{\gamma[K(r-\mu+\epsilon)-rO]}\equiv B(O).
\end{equation}

One can solve $S$ from \eqref{9}:
\begin{equation}\label{12}
    S\equiv S(O,B)=\frac{C}{\beta}g e^{-FO/Cg},
\end{equation}
where $g$ depends on $O+B$ and $F$ depends on $g$. Now the substitution of \eqref{14} and \eqref{12} into \eqref{10} results in an implicit equation of $O$ only:
\begin{equation}\label{16}
    f\left(\frac{O}{2}+B(O)-S(O,B(O))\right)=\frac{\epsilon K}{K(r-\mu+\epsilon)-rO}.
\end{equation}
Hence for a fixed set of parameters, any root $O_*>0$ of \eqref{16} corresponds to an equilibrium point $(O_*,B(O_*),S(O_*,B(O_*)))$ of \eqref{7}-\eqref{9}. While direct analysis of \eqref{16} is not simple due to the complicated definitions of $O(B)$ and $S(O,B)$, numerical calculation of \eqref{16} is relatively easy with a symbolic mathematics software.

We define the functions on the left and right hand side of \eqref{16} to be
\begin{align}\label{21}
    L(O)&=f\left(\frac{O}{2}+B(O)-S(O,B(O))\right),\\
    R(O)&=\frac{\epsilon K}{K(r-\mu+\epsilon)-rO}.
\end{align}
From \eqref{16}, intersection points of the graphs of $L(O)$ and $R(O)$ are equilibrium points. We observe that $L(O)$ is bounded by
$1$ and $R(O)$ is unbounded as $O\to K_*=K(r-\mu+\epsilon)/r$. Thus if $L(0)>R(0)$, then \eqref{16} has at least one root with positive $O$ from the intermediate-value theorem; and if $L(0)<R(0)$, then \eqref{16} may have no or two zeros for most cases, or one zero in the case that $L(O)$ and $R(O)$ are tangent to each other. From \eqref{7} we see that
\begin{equation*}
    L(0)=f(-C/\beta), \;\;\; \text{ and } \;\; R(0)=\frac{\epsilon}{r-\mu+\epsilon}.
\end{equation*}

From a different point of view, one can consider the equations \eqref{7}-\eqref{9} with a bifurcation analysis and linearization. Linearizing \eqref{7}-\eqref{9} at the trivial equilibrium $(O,B,S)=(0,0,C/\beta)$, we obtain the Jacobian matrix to be
\begin{equation}\label{20}
   J(0,0,C/\beta)= \left(
  \begin{array}{ccc}
    f(-C/\beta)(r-\mu+\epsilon)-\epsilon & 0 & 0 \\
    f(-C/\beta)(\mu-\epsilon)+\epsilon & -\gamma & 0 \\
    C g'(0) - F(C) & C g'(0) & -\beta \\
  \end{array}
\right).
\end{equation}
Since the Jacobian matrix $J(0,0,C/\beta)$ is lower triangular, the three diagonal entries are eigenvalues.

We discuss the equilibrium problem in several cases:

\noi \underline{Case 1}: If $r\le \mu$ (the birth rate smaller than the natural death rate),  then the trivial one $(0,0,C/\beta)$ is the only equilibrium. In this case $O'<0$, thus the live oysters are destined to go extinct. So the equilibrium $(O,B,S)=(0,0,C/\beta)$ is globally asymptotically stable.

\noi \underline{Case 2}: If $\mu<r\le \mu+\epsilon$ (the birth rate larger than the natural death rate, but smaller than the combined death rate due to natural cause and due to sediment), positive equilibrium points are possible. We notice that $f(0)=1/2$ from  \eqref{5}, so $r<\mu+\epsilon$ is equivalent to
\begin{equation*}
    f(0)(r-\mu+\epsilon)\le \epsilon.
\end{equation*}
This implies that for any $C\ge 0$, $f(-C/\beta)(r-\mu+\epsilon)-\epsilon<f(0)(r-\mu+\epsilon)-\epsilon<0$ since $f$ is an increasing function. Thus the trivial equilibrium $(0,0,C/\beta)$ is locally stable for any $C\ge 0$.

For a critical value $r_*\in (\mu,\mu+\epsilon)$, if $r>r_*$, then equation \eqref{16} has exactly two positive roots when $C=0$ if $f(d)$ is a concave function for $d>0$. And for a fixed $r$ value satisfying $r_*<r<\mu+\epsilon$, there exists another critical value $C^*(r)>0$ such that equation \eqref{16} has exactly two positive roots for $0\le C< C^*(r)$. When the maximum sediment deposition rate $C$ is large, the system \eqref{1}-\eqref{3} can only have the trivial equilibrium. Hence the parameter region for existence of two positive equilibria when  $\mu<r\le\mu+\epsilon$  is $\{(r,C): r_*<r\le\mu+\epsilon, 0<C<C^*(r)\}$, given that all other parameters are fixed.

\noi \underline{Case 3}: If $r>\mu+\epsilon$ (the birth rate larger than the combined death rate due to predation, disease and sediment), then
\begin{equation*}
    f(0)(r-\mu+\epsilon)>\epsilon.
\end{equation*}
The monotonicity of $f$ assumed in \eqref{5} implies that there exists a unique $C_*(r)>0$ such that $f(-C/\beta)(r-\mu+\epsilon)-\epsilon>0$ for $C>C_*(r)$, and
$f(-C/\beta)(r-\mu+\epsilon)-\epsilon<0$ for $C<C_*(r)$. Thus in this case, the equilibrium $(0,0,C/\beta)$ is locally stable when $C>C_*(r)$, and it is unstable when
$0<C<C_*(r)$.

\begin{figure}[h]
\centering
\begin{picture}(450,100)
\linethickness{2.5pt}
\put(200,0){\line(-1,0){100}}
\qbezier(100,0)(100,80)(0,80)
\put(450,0){\line(-1,0){100}}
\qbezier(420,40)(420,60)(350,80)
\linethickness{1pt}
\put(100,0){\line(-1,0){100}}
\put(350,0){\line(-1,0){100}}
\qbezier(350,0)(420,20)(420,40)
\end{picture}
\caption{\label{g:0}Schematic Diagrams of the Forward and Backward Bifurcations. The horizontal direction is the parameter, and the vertical direction is the state variable. Stable equilibria: thick curves; unstable equilibria: thin curves. (left) forward bifurcation; (right) backward bifurcation.}
\end{figure}
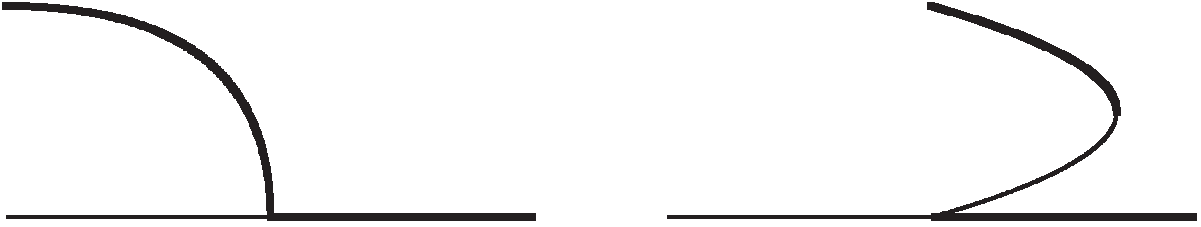

The critical value $C=C_*(r)$ is a bifurcation point where a branch of nontrivial equilibrium points emanates from the line of trivial equilibria $(C,O,B,S)=(C,0,0,C/\beta)$. The bifurcation is  backward if the bifurcating equilibria are unstable, otherwise it is forward: the bifurcating equilibria are stable (Fig.~\ref{g:0}). When the bifurcation is forward,  a unique equilibrium exists for $0<C<C_*(r)$, and there is no nontrivial equilibrium for $C>C_*(r)$. Conversely, if the bifurcation is backward, then there is a range of values of $C>C_*(r)$ for which the system has two nontrivial equilibria. If the sediment volume is too large, there is a largest value $C=C^*(r)$ such that the system only has the trivial equilibrium when $C>C^*(r)$. The parameter region for two positive equilibria occurs when  $r>\mu+\epsilon$  is $\{(r,C): r>\mu+\epsilon, C_*(r)<C<C^*(r)\}$, given that all other parameters are fixed. Note that the bifurcation occurring  at $C=C_(r)$ is a transcritical one. The terms \lq\lq backward\rq\rq\ and \lq\lq forward\rq\rq\ are about the branch of positive equilibria,  and they are similar to the ones used in epidemic models (see \citep{Hadeler97}.)

In the Appendix, we show that the direction of the branch of bifurcating positive equilibria is determined by
\begin{equation}\label{I}
    I=
\frac{\la
(r-\mu+\epsilon)K}{2r}-\frac{f(-C_*/\beta)}{f'(-C_*/\beta)}
+\ds\frac{\la \epsilon K}{\gamma}-\frac{\epsilon K}{\gamma \beta}\left(C_* g'(0) - F(C_*)+\frac{C_* g'(0)\epsilon r}{\gamma(r-\mu+\epsilon)}\right).
\end{equation}
If $I<0$ then the bifurcation is forward; if $I>0$ then it is backward which implies bistable parameter ranges.

Therefore the question of bistability for this case is reduced to whether $I>0$. Notice that $g'(0)<0$ and $F(C_*)>0$, hence
\begin{equation*}
   I_1= \frac{\la
(r-\mu+\epsilon)K}{2r}
+\ds\frac{\la \epsilon K}{\gamma}-\frac{\epsilon K}{\gamma \beta}\left(C_* g'(0) - F(C_*)+\frac{C_* g'(0)\epsilon r}{\gamma(r-\mu+\epsilon)}\right)>0,
\end{equation*}
and the positivity of $I=I_1-I_2$ depends on the competition between $I_1$ and $I_2=\ds\frac{f(-C_*/\beta)}{f'(-C_*/\beta)}>0$. Here $I_2$ only depends on the form of $f$ and $C_*/\beta$, while $I_1$ depends on many other parameters. Notice that $C_*$ is determined by $f$ and $\ds\frac{r-\mu+\epsilon}{\epsilon}$. If we fix the values of $r,\mu,\epsilon$, and $\beta$, then one can increase $I_1$ to generate bistability by (i) increasing carrying capacity $K$; (ii) decreasing the oyster shell degradation rate $\gamma$; (iii) increasing $|g'(0)|$, the decay rate of sediment deposition with reef height; or (iv) increasing $F(C_*)$, which represents oyster filtration efficiency.

\begin{figure}[h]
\centering
  \includegraphics[width=0.5\textwidth]{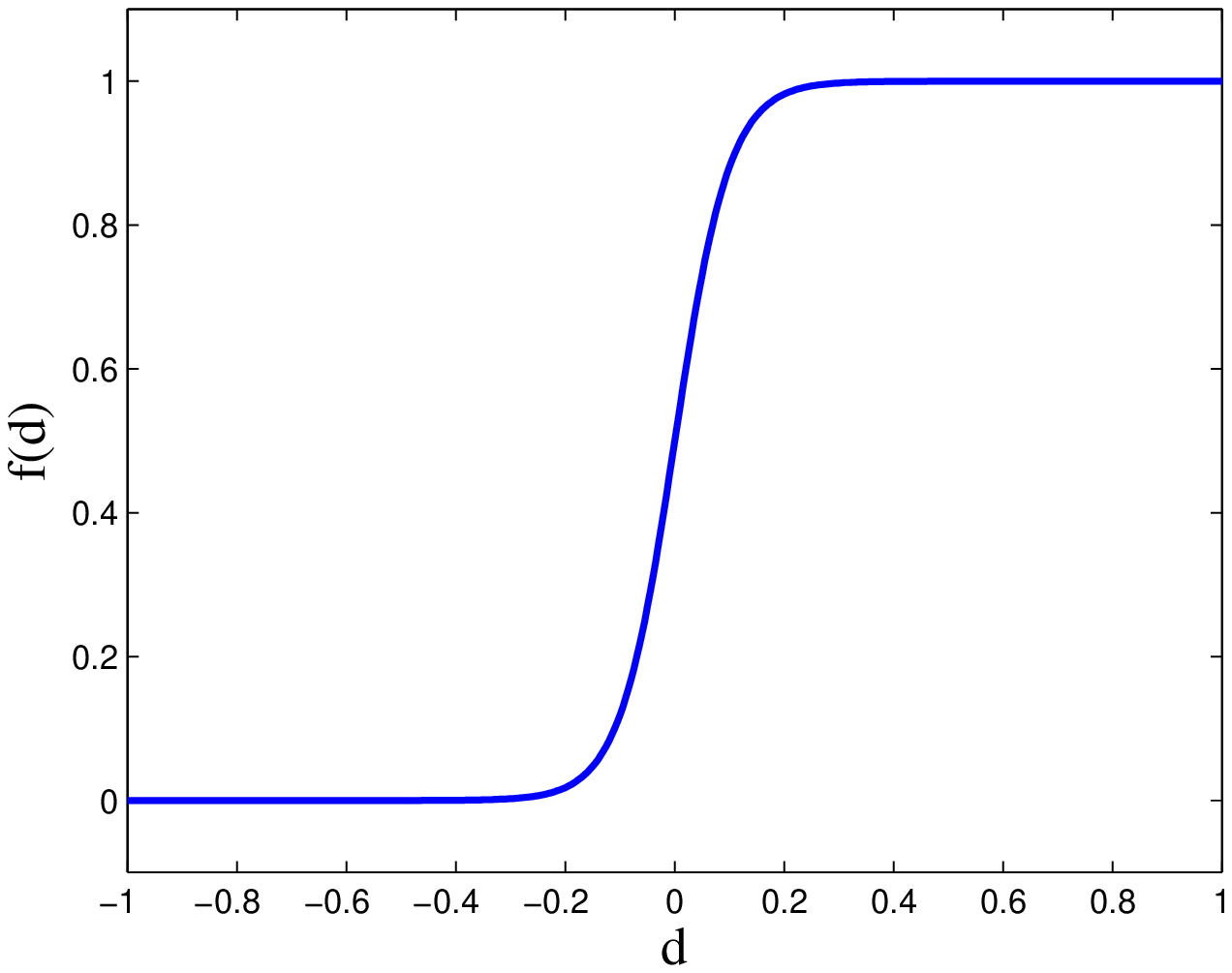}\includegraphics[width=0.5\textwidth]{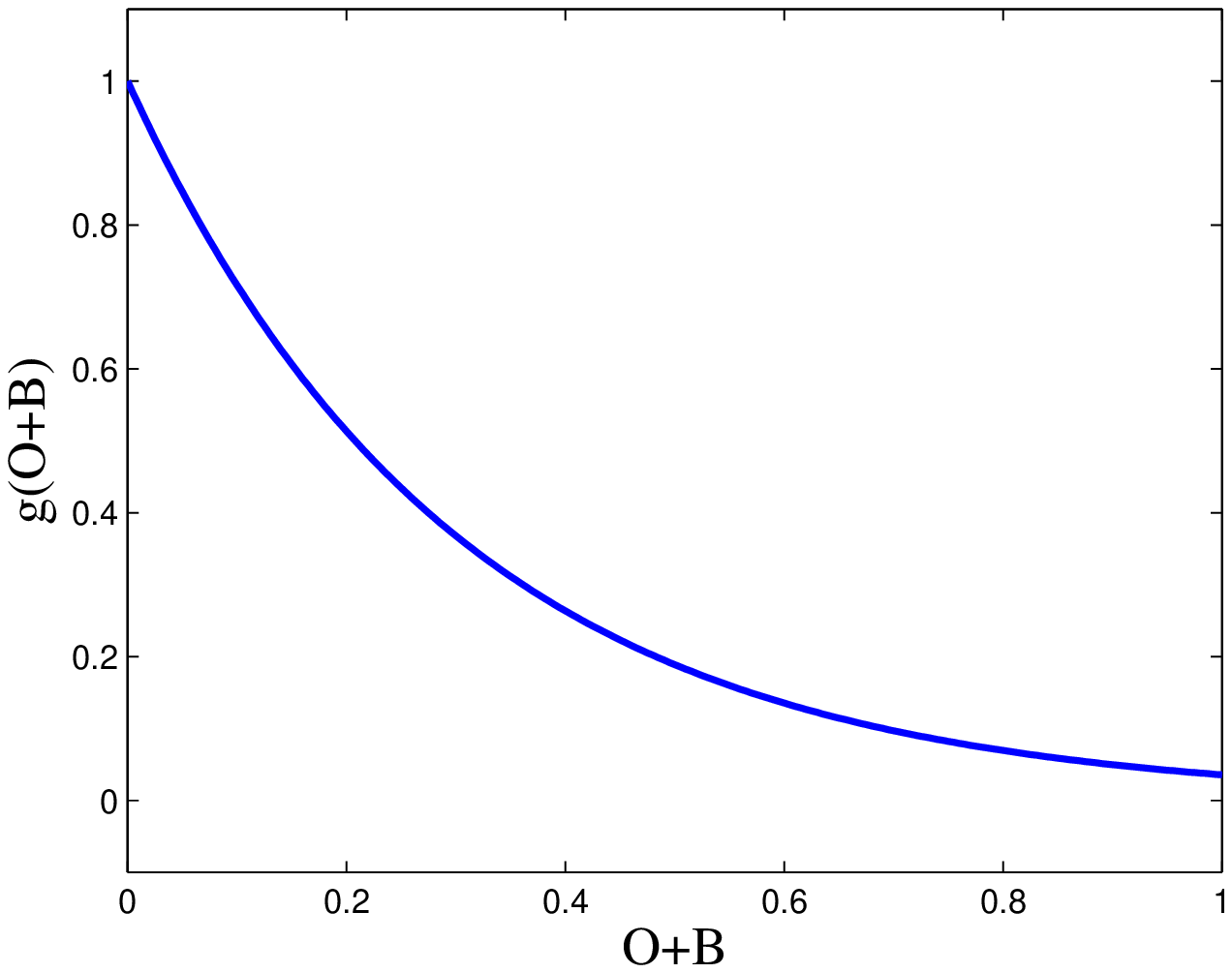}
  \includegraphics[width=0.5\textwidth]{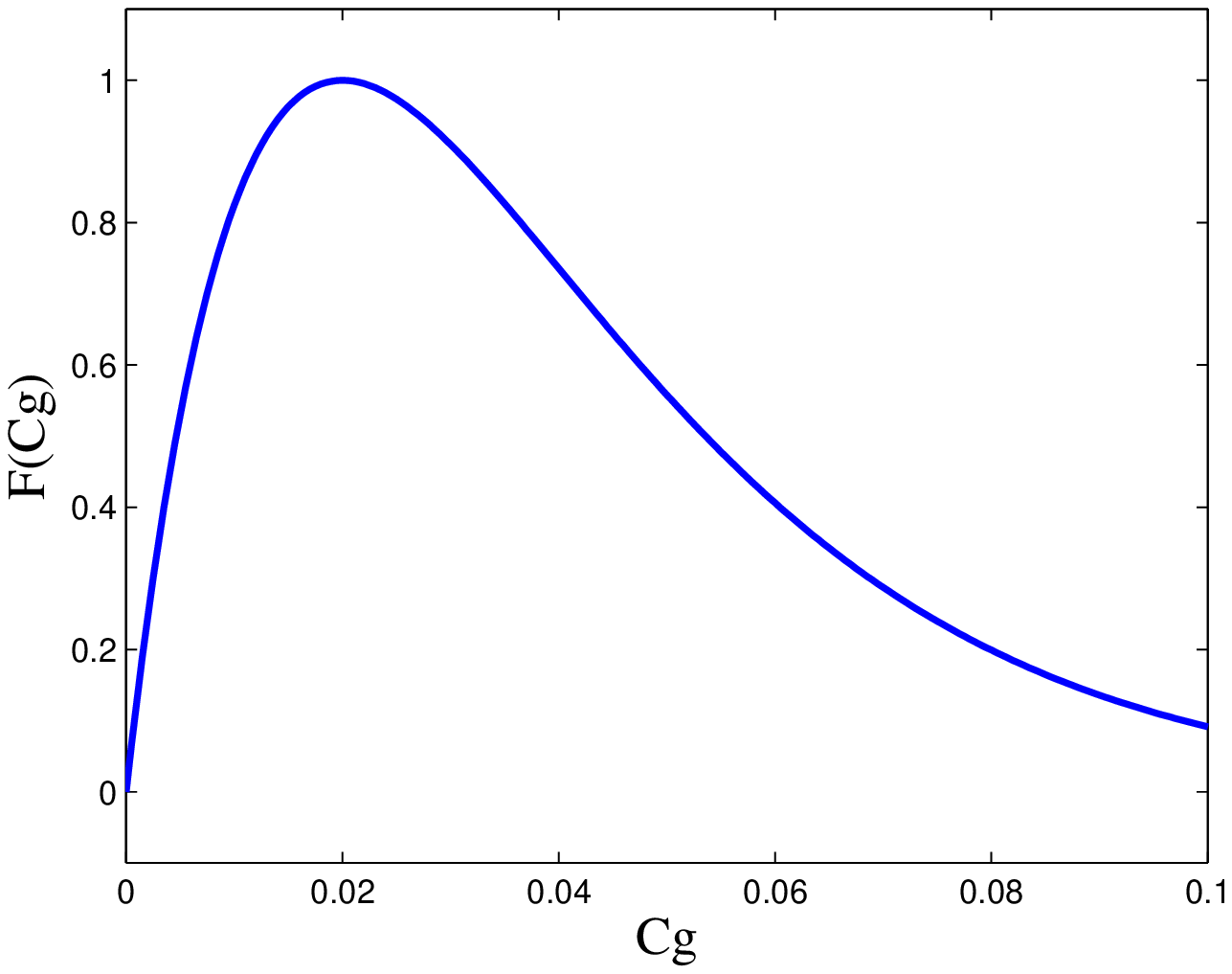}
  \caption{\label{g:1} Graphs of functions $f(d)$, $g(x)$ and $F(y)$ defined in \eqref{333}. (left) $f(d)$ with $h=20$; (middle) $g(x)$ with $\eta=10/3$; (right) $F(y)$ with $y_0=0.02$ and $F_0=1$.}
\end{figure}

\subsection{Numerical simulations}
\label{sec4b}

To illustrate our results numerically, we define the functions $f(d)$, $g(x)$, and $F(y)$, which satisfy the desired conditions \eqref{5}, \eqref{g}, \eqref{F}:
\begin{equation}\label{333}
    \begin{split}
       \ds f(d) &= \frac{1}{1+e^{-h d}}, \;\; d=\frac{ O}{2}+B-S, \\
        \ds g(x)& =e^{-\eta x}, \;\;\;\; (g(O+B)=e^{-\eta (O+B)}),\\
        \ds F(y)&=\frac{F_0}{y_0}ye^{(y_0-y)/y_0}, \;\;\;\; (F(Cg)=\frac{F_0 Cg}{y_0}e^{(y_0-Cg)/y_0}).
    \end{split}
\end{equation}
Here $h>0$ is a parameter that adjusts the shape of the sigmoid function $f$. For larger $h$, the function $f$ has a narrower transition where the function value jumps from $0$ to $1$.  In the definition of $g$, $\eta$ is the decay rate of the exponential function. The per volume filtration rate $F=F(Cg)$ is a function of $Cg$ and is of the Ricker type, where $F_0$ represents the maximum filtration rate, achieved at $y=y_0$ (Fig.~\ref{g:1}).

With the nonlinear functions $f$, $g$ and $F$ defined as in \eqref{333}, we assume the following set of parameters:
\begin{table}[H]
\centering
\begin{tabular}{|c|c|c|c|c|c|c|c|c|c|c|c|}
  \hline
  Parameter  & $r$ & $K$ & $\mu$ & $\epsilon$ & $\gamma$ & $\eta$ & $y_0$ & $F_0$ & $\beta$ & $h$ & $C$ \\
  \hline
  Value  & $1$ & $0.3$ & $0.4$ & $0.94$ & $0.7$ & $3.33$ & $0.02$ & $1$ & $0.01$ & $20$ & $0.02$ \\
  \hline
\end{tabular}
\caption{A sample set of reasonable parameters\label{table1}}
\end{table}

\begin{figure}[h]
\centering
  \includegraphics[width=0.5\textwidth]{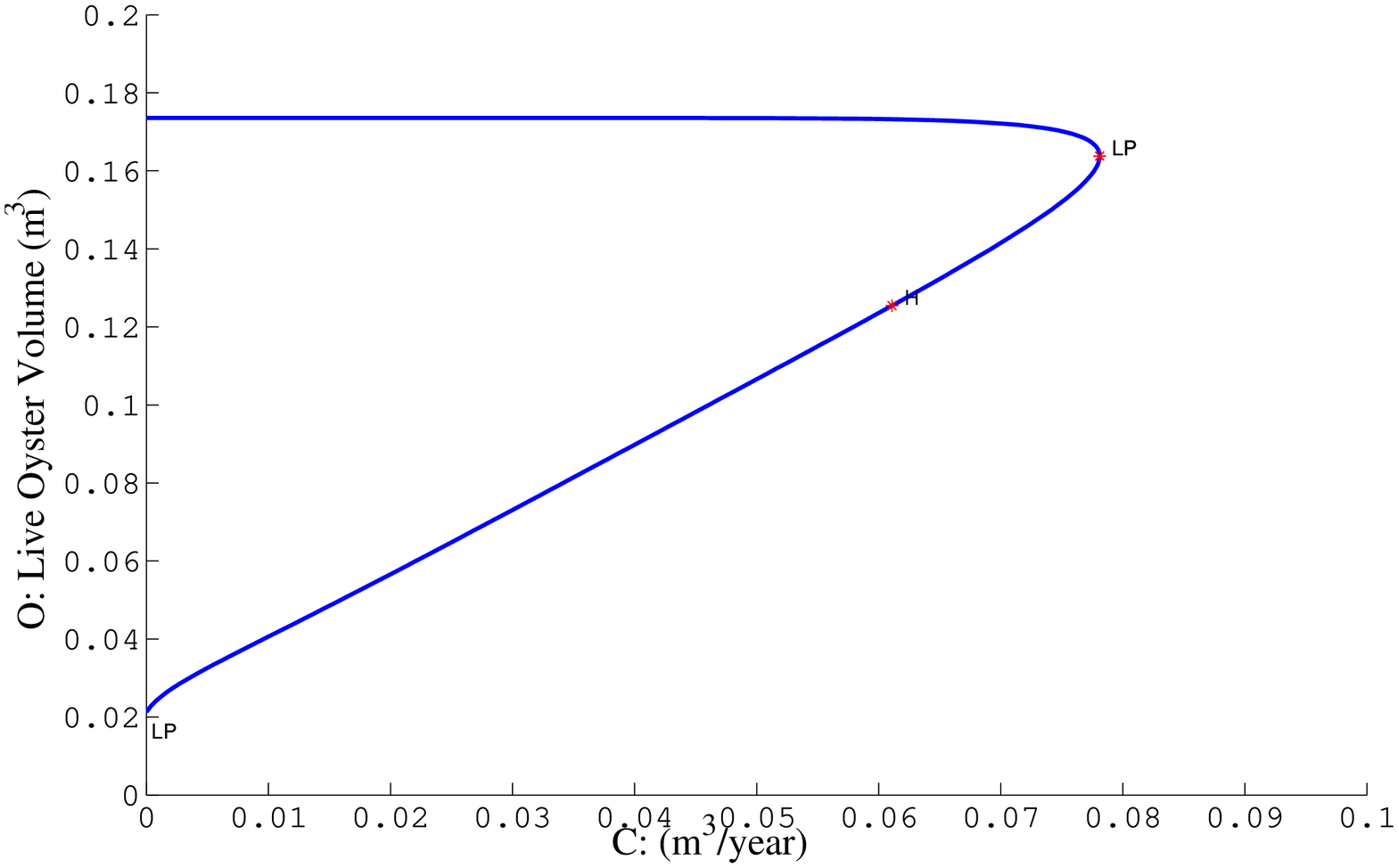}\includegraphics[width=0.5\textwidth]{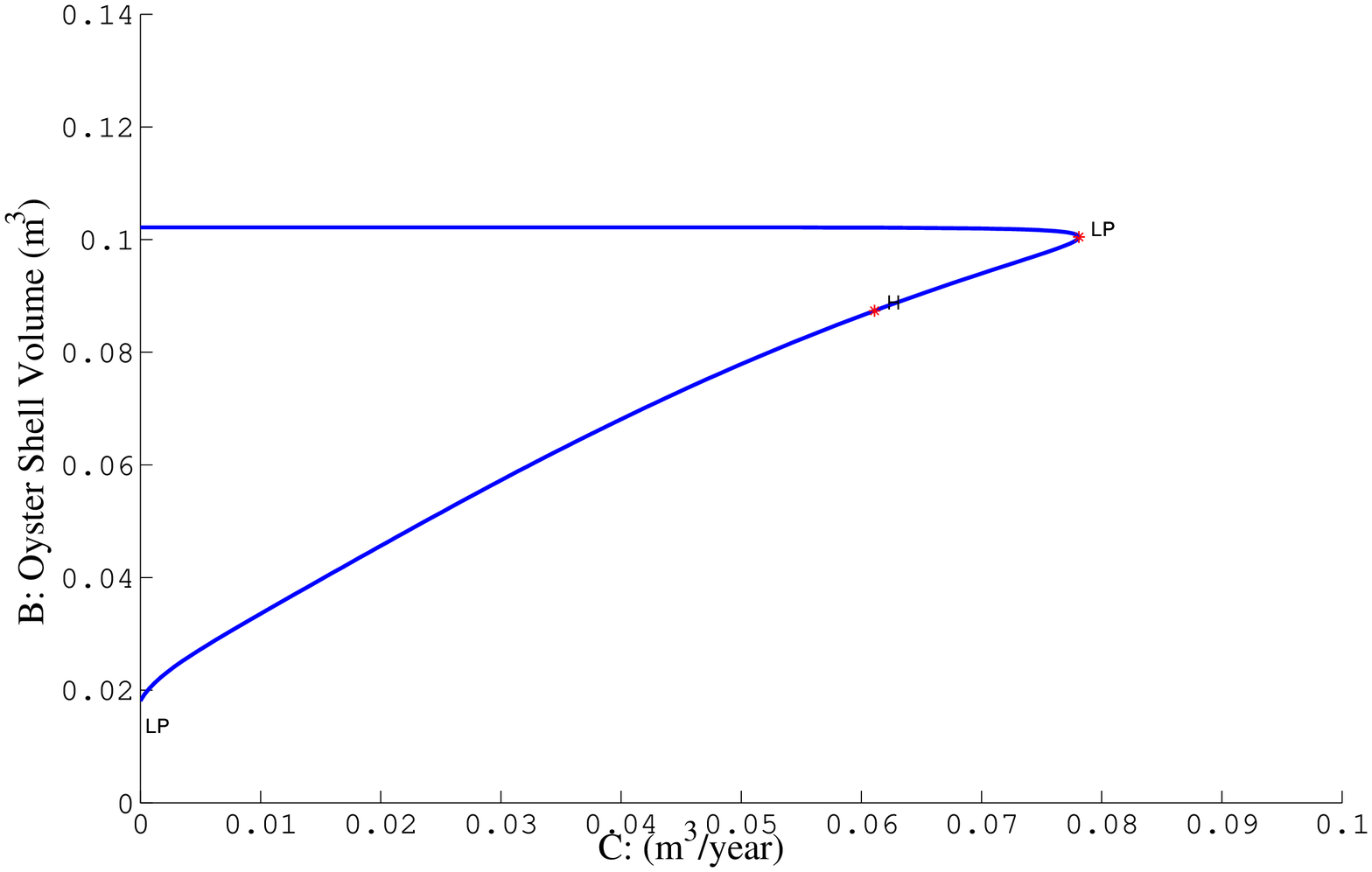}
  \includegraphics[width=0.5\textwidth]{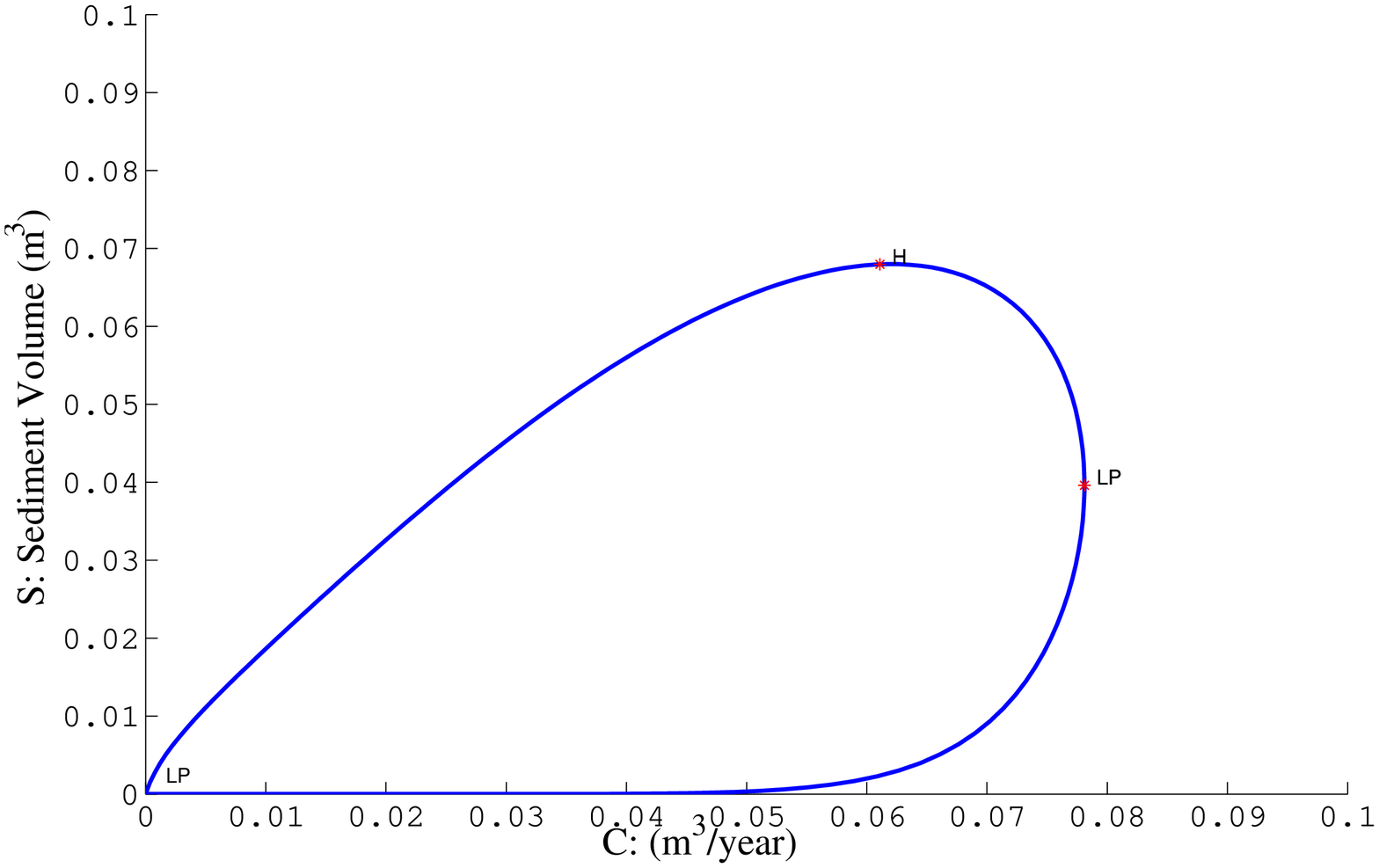}
  \caption{\label{g:b1} Bifurcation diagram of \eqref{7}-\eqref{9}, with parameters given in Table \ref{table1}. Here the horizontal axis is sediment deposition rate $C$, and the vertical axis is (upper left) live oyster volume $O$, (upper right) dead shell volume $B$ and (lower) deposited sediment $S$. The point labeled \lq LP\rq\ in the interior is the saddle-node bifurcation point, and the point labeled \lq H\rq\ is a neutral saddle point. The bifurcation diagram is generated by \texttt{Matlab} package \texttt{MatCont}.}
\end{figure}

For the parameters given by Table \ref{table1}, there are two positive  equilibrium points $(O_1, B_1, S_1)=(0.0566,0.0456,0.0326)$ and $(O_2, B_2, S_2)=(0.1736,0.1022,1.0645\times 10^{-7})$; thus, the parameter set in Table \ref{table1} is in the bistable region. Freeing the parameter $C$ gives a bifurcation diagram (see Fig.~\ref{g:b1}) with two positive  equilibria for all $0<C<C^*$, where $C^*\approx 0.078$ is a saddle-node bifurcation point  where the curve bends back. This bifurcation diagram confirms the description we give in Section \ref{sec4} Case 2, as $0.4=\mu<r=1<\mu+\epsilon=1.34$.

\begin{figure}[t]
\centering
  \includegraphics[width=0.5\textwidth]{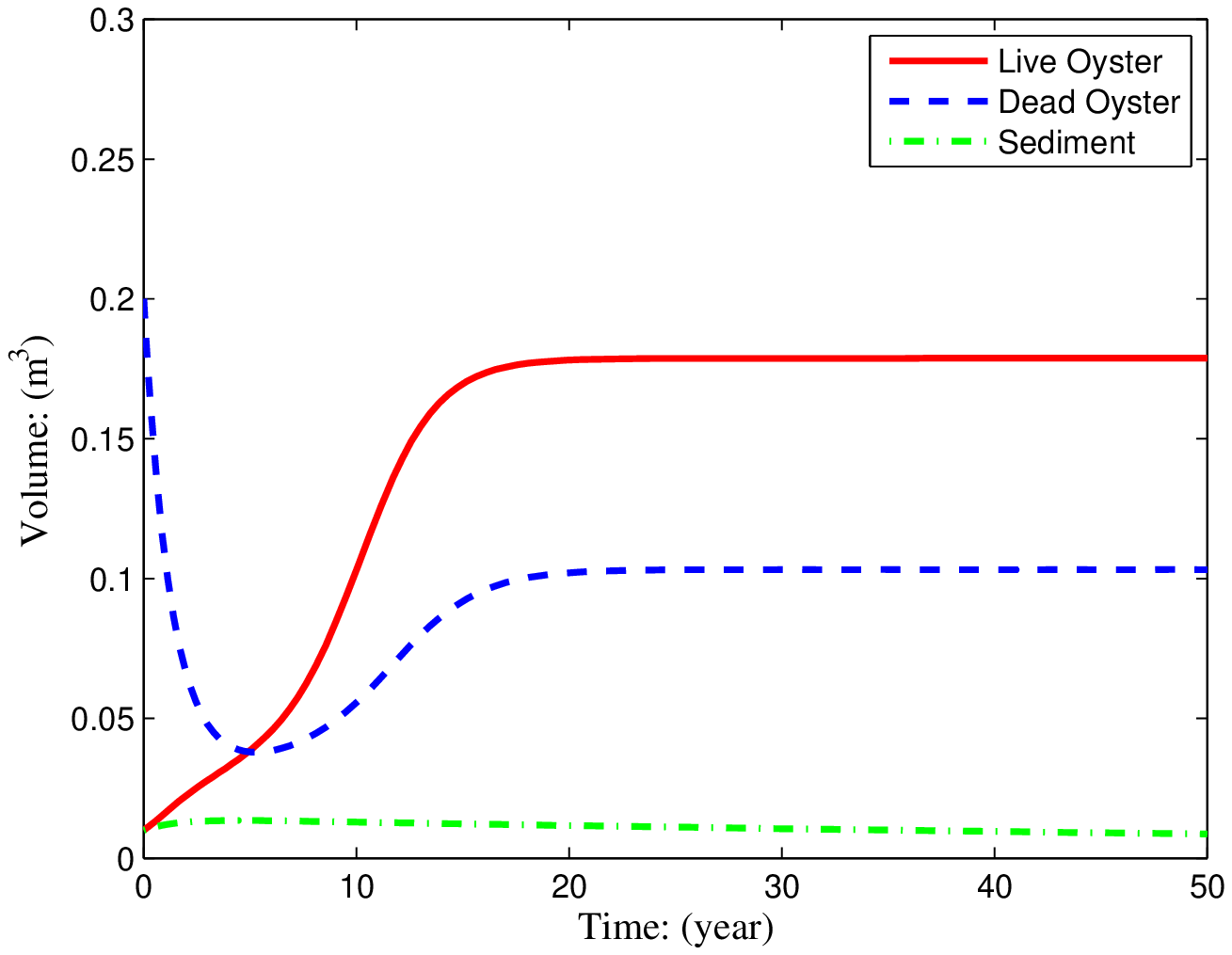}\includegraphics[width=0.5\textwidth]{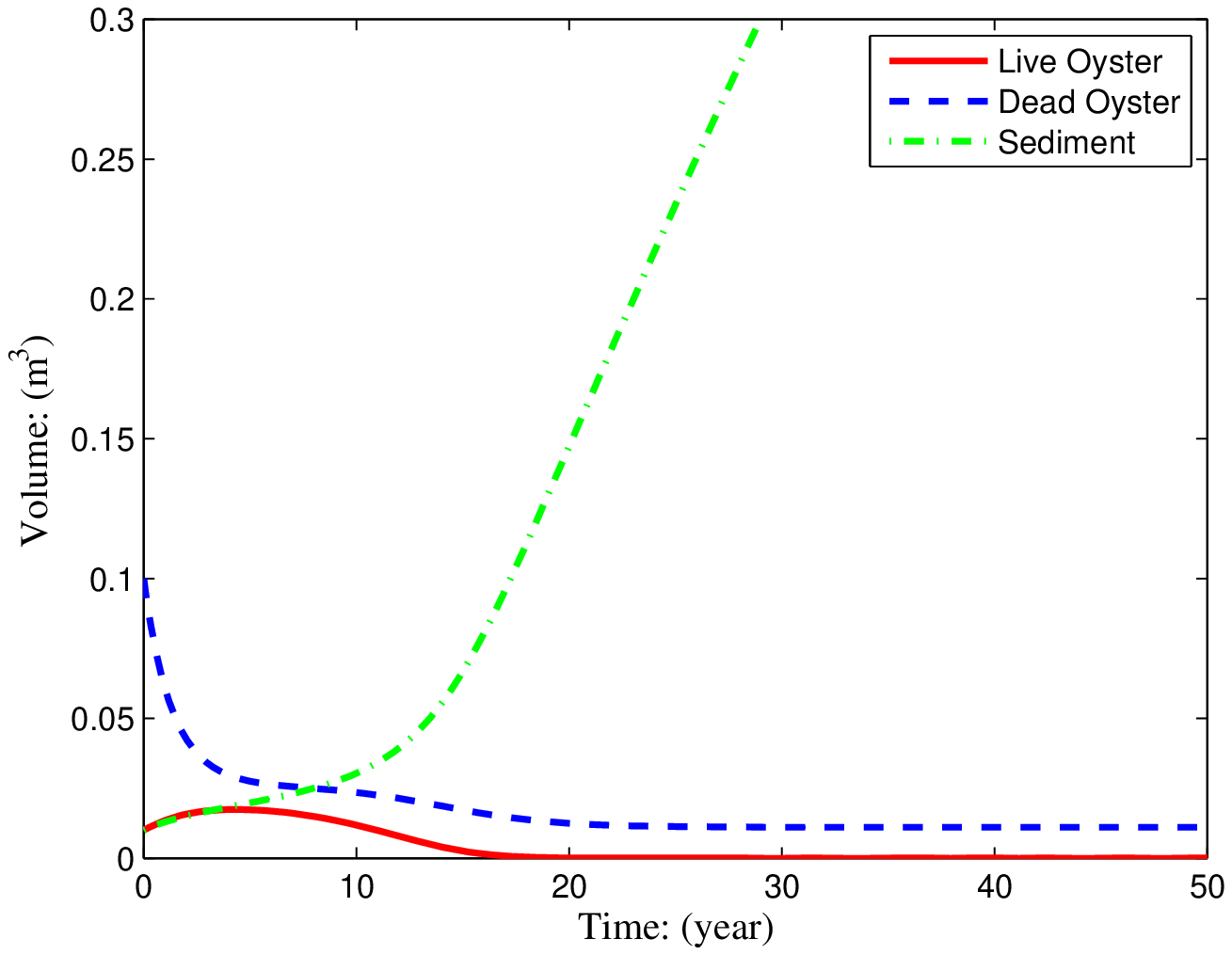}
  \includegraphics[width=0.5\textwidth]{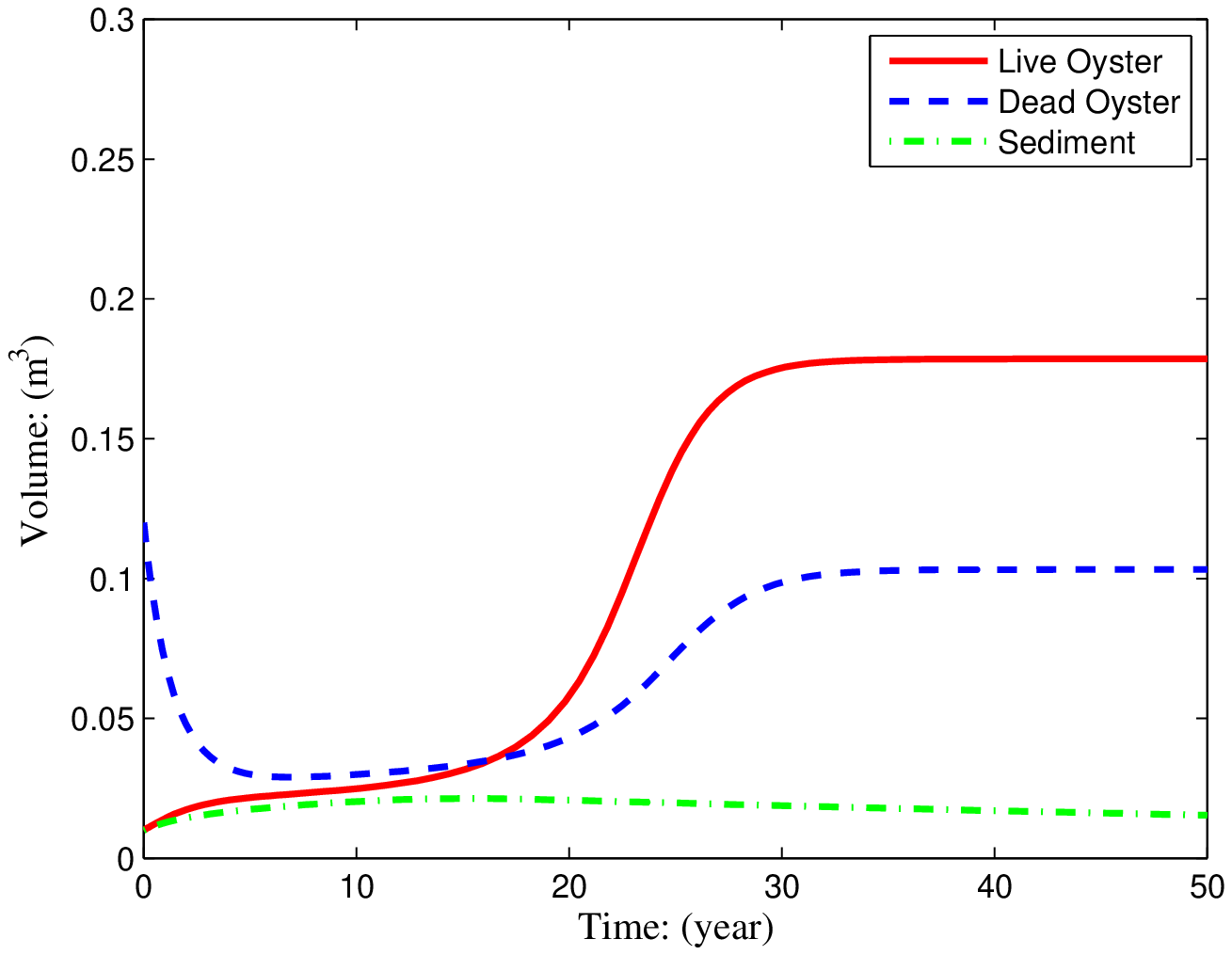}\includegraphics[width=0.5\textwidth]{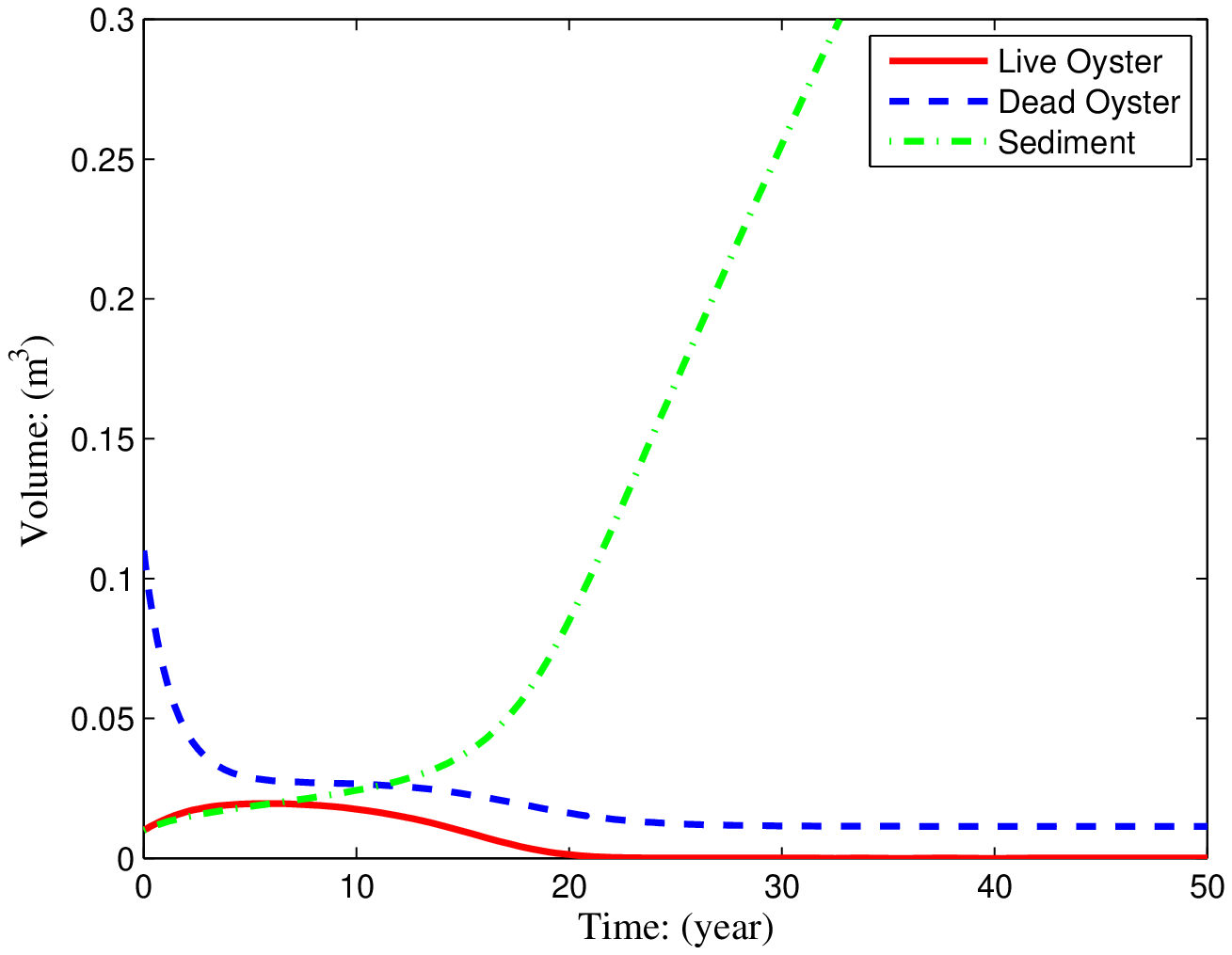}
  \caption{\label{g:b2} Numerical solution ($0\le t\le 50$) of \eqref{7}-\eqref{9} with parameters given in Table \ref{table1} for various initial reef heights. For all cases $O(0)=0.01$ and $S(0)=0.01$. (upper left)  $B(0)=0.20$; (upper right) $B(0)=0.10$; (lower left) $B(0)=0.12$; (lower right) $B(0)=0.11$.}
\end{figure}

\begin{figure}[h]
\centering
  \includegraphics[width=0.5\textwidth]{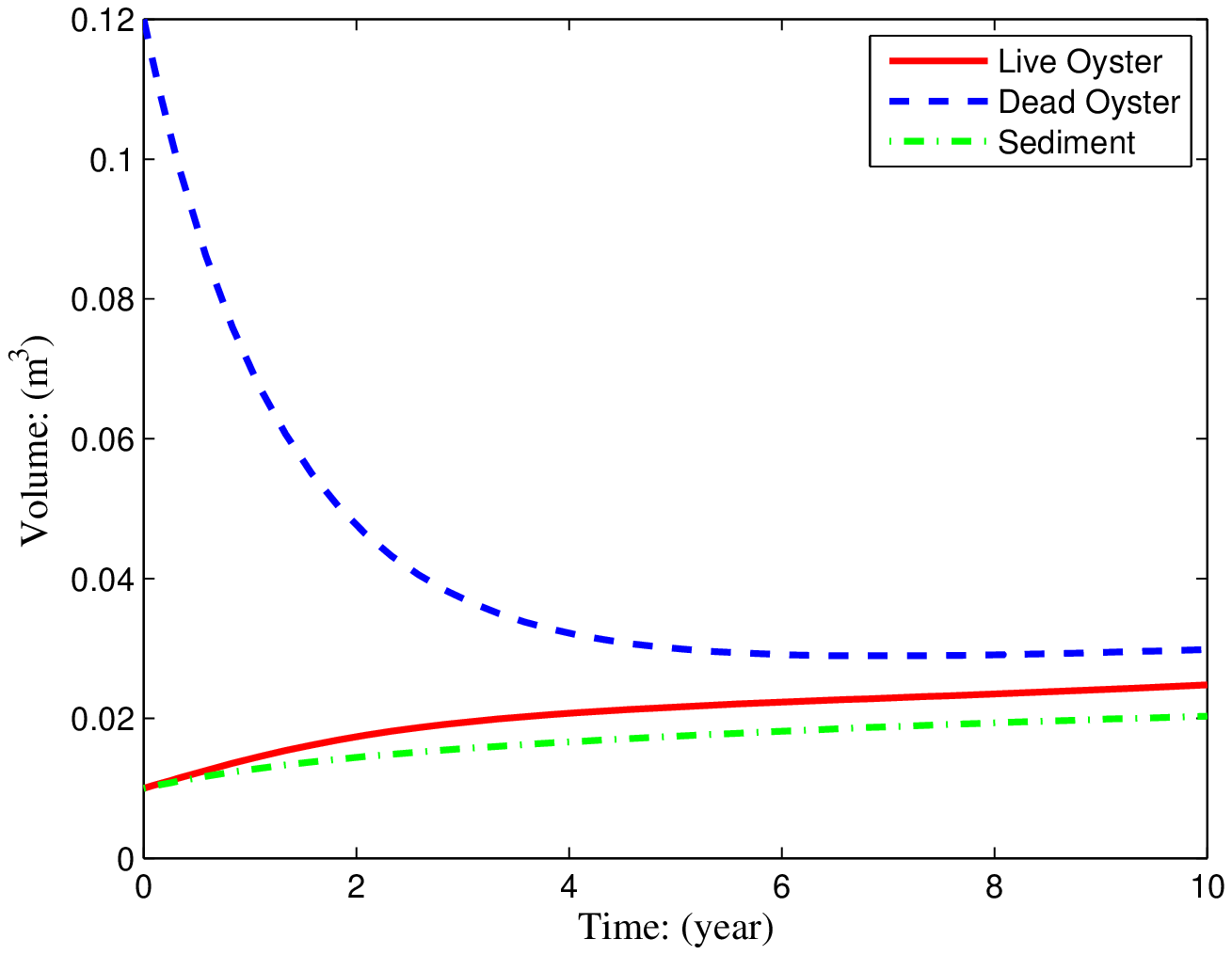}\includegraphics[width=0.5\textwidth]{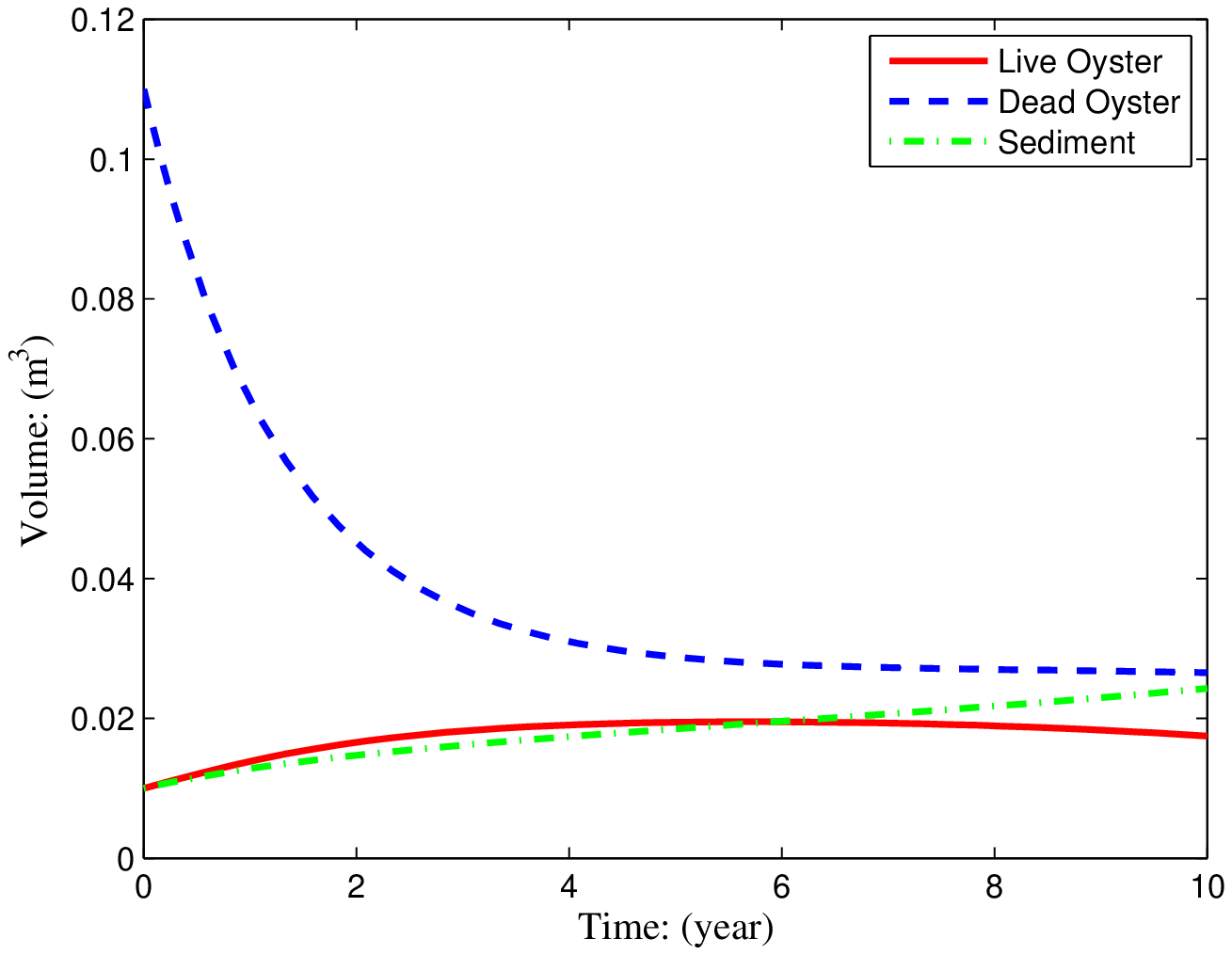}
  \caption{\label{g:b4} Numerical solution ($0\le t\le 10$) of \eqref{7}-\eqref{9} with parameters given in Table \ref{table1} for various initial reef heights.
  (left) Initial value $O(0)=0.01$, $B(0)=0.12$ and $S(0)=0.01$; (right) Initial value $O(0)=0.01$, $B(0)=0.11$ and $S(0)=0.01$}
\end{figure}

We use numerical simulation to examine the bistable dynamics of \eqref{7}-\eqref{9} with parameters given in Table \ref{table1}. We use the initial value of $O(0)=0.01$ and $S(0)=0.01$; \textit{i.e.}, there is a small amount of live oyster and also a small amount of sediment initially. We chose several different values of  $B(0)$: $B(0)=0.20$, $0.10$, $0.12$ and $0.11$ (Fig.~\ref{g:b2}). For larger $B(0)$, the  oyster population survives and reaches the large stable equilibrium $(O_1, B_1, S_1)$, whereas the smaller $B(0)$ will drive the oyster population to local extinction. The critical level of initial reef height $B(0)$ is between $0.11$ and $0.12$. The transient dynamics with $B(0)=0.12$ and $B(0)=0.11$ when $0\le t\le 10$ (Fig.~\ref{g:b4}) demonstrates that at the higher reef height, live oyster volume increases and eventually curbs sediment volume to a very small level due to oyster filtration and reef height. The slightly lower reef does not permit live oysters to filter the sediment sufficiently, and the sediment eventually covers both live and dead oysters. The specific reef heights (\textit{i.e.}, $0.11$ and $0.12$) that discriminate the two trajectories towards stability are likely to differ depending on natural variation in other parameters of the three equations, such that they should not be viewed as rigid values under all conditions. Rather, the key point is that a slight shift in initial conditions can drive the system towards two dramatically different trajectories.

\begin{figure}[h]
\centering
  \includegraphics[width=0.5\textwidth]{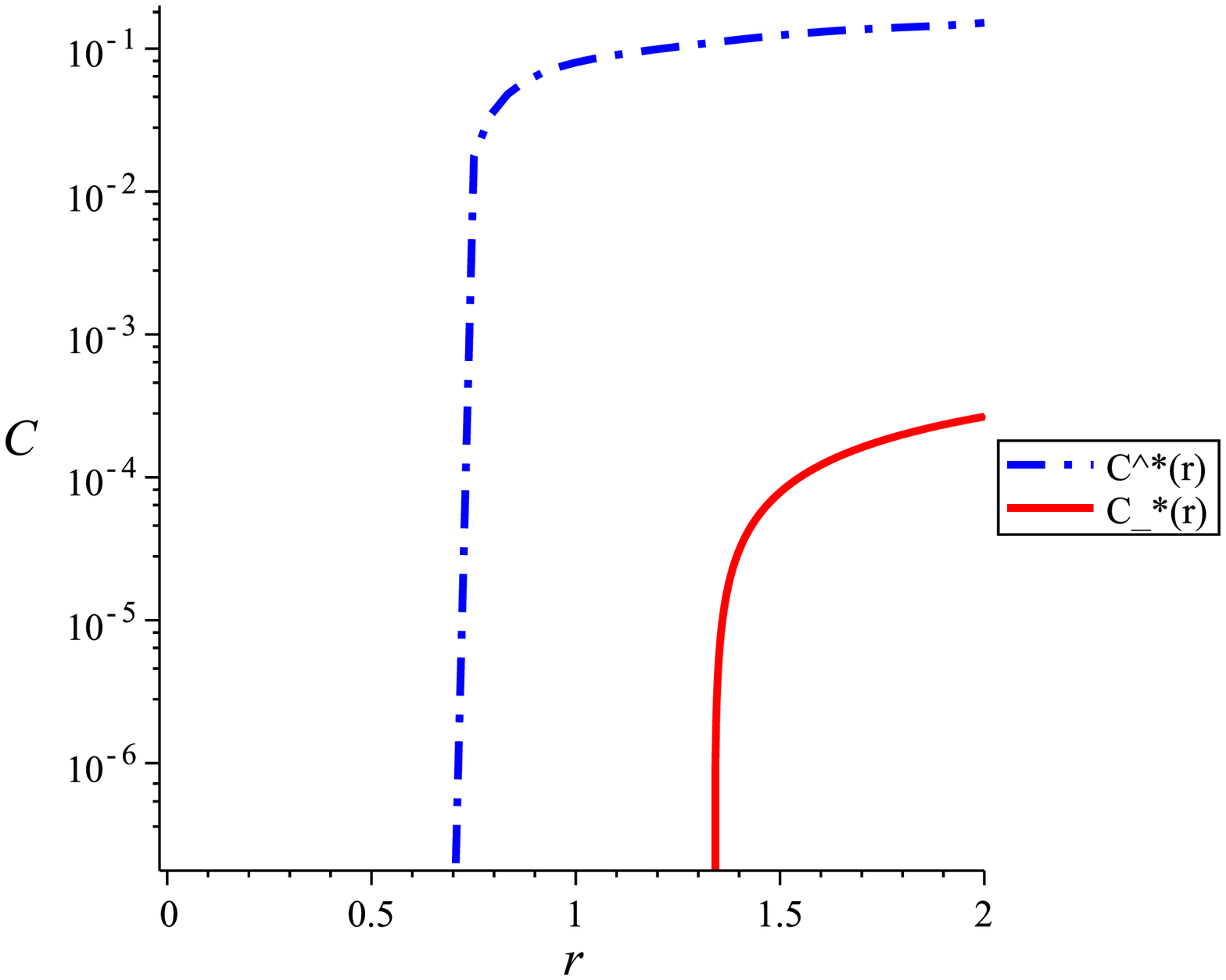}\includegraphics[width=0.5\textwidth]{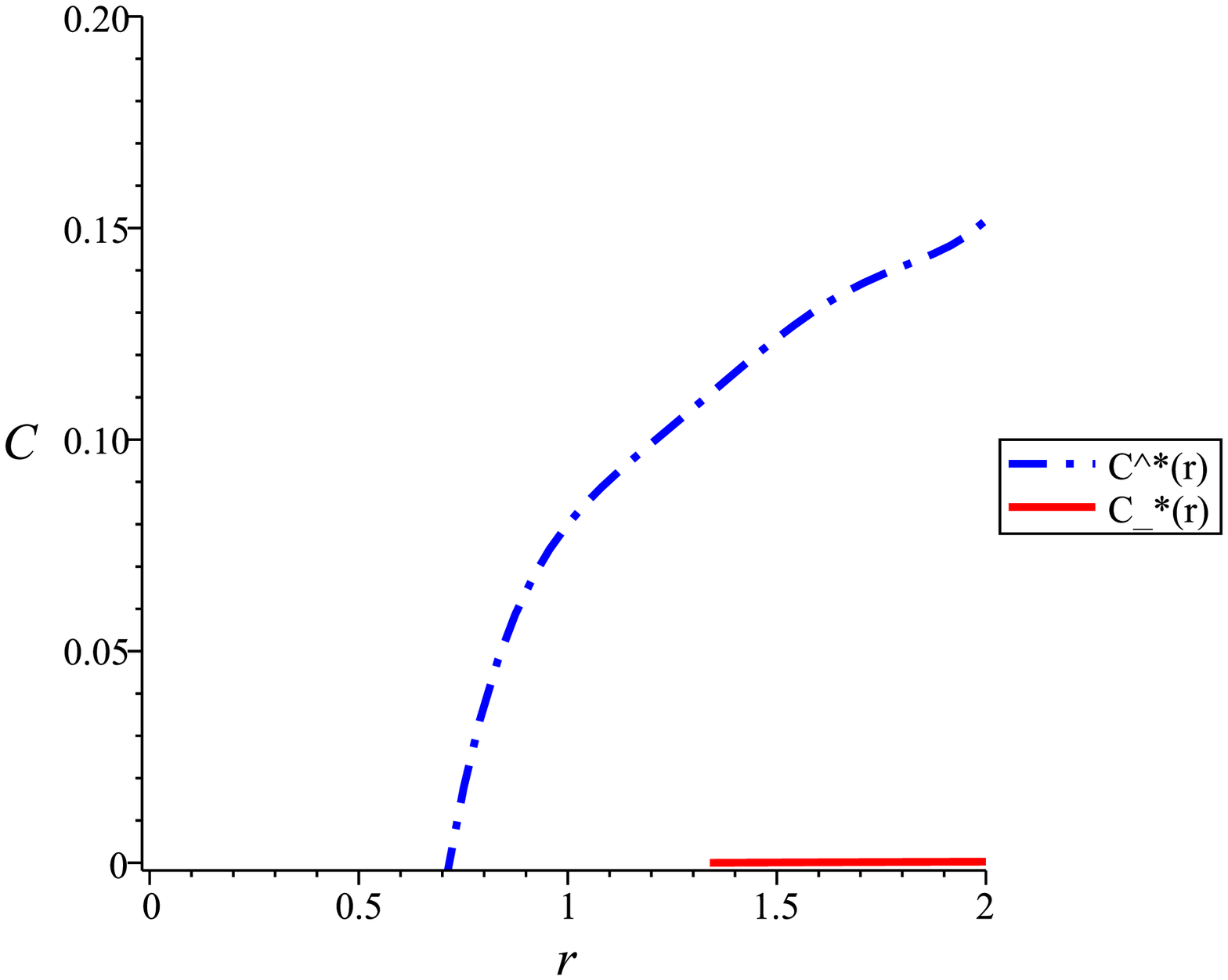}
   \caption{\label{g:b5} The graph of $C^*(r)$ (curve of saddle-node bifurcation points) and $C_*(r)$ (curve of transcritical bifurcation points), with parameters given in Table \ref{table1} except $r$ and $C$.  Here the horizontal axis is oyster growth rate $r$, and the vertical axis is (left) $\log(C)$, (right) $C$. }
\end{figure}

The bistability displayed with parameter values given in Table \ref{table1} is not anomalous (Section \ref{sec4}); the parameter range for bistability is robust. For example, given all other parameter values as in Table \ref{table1} except the oyster growth rate $r$ and sediment deposition rate $C$, then the range of parameters $(r,C)$ to produce bistability has been shown to be
\begin{equation*}
    \{(r,C): r_*<r<\mu+\epsilon, 0<C<C^*(r)\} \;\;\text{ and }\;\; \{(r,C): r>\mu+\epsilon, C_*(r)<C<C^*(r)\}.
\end{equation*}
The graphs of $C^*(r)$ (Fig. \ref{g:b5}, upper curve) and $C_*(r)$ (Fig. \ref{g:b5}, lower curve) are monotonically increasing functions of $r$. The first critical value $r=r_*\approx 0.739$ is where the curve $C=C^*(r)$ emerges from $C=0$, and the second critical value $r=\mu+\epsilon=1.34$. The region between the two curves is the \lq\lq bistable regime\rq\rq, and the chosen parameter value $(r,C)=(1,0.02)$ is in that region. The region above the bistable one is the \lq\lq extinction regime\rq\rq\ where the trivial equilibrium $(0,0,C/\beta)$ is globally asymptotically stable; the region below the bistable one (which is very small and can only be detected on the log-plot) is the \lq\lq persistent regime\rq\rq\ where the oysters will persist irrespective of initial reef height. Note that  the estimated growth rate range $0.7\le r\le 1.3$ is dominated by the bistable regime.

\begin{figure}[h]
\centering
  \includegraphics[width=0.5\textwidth]{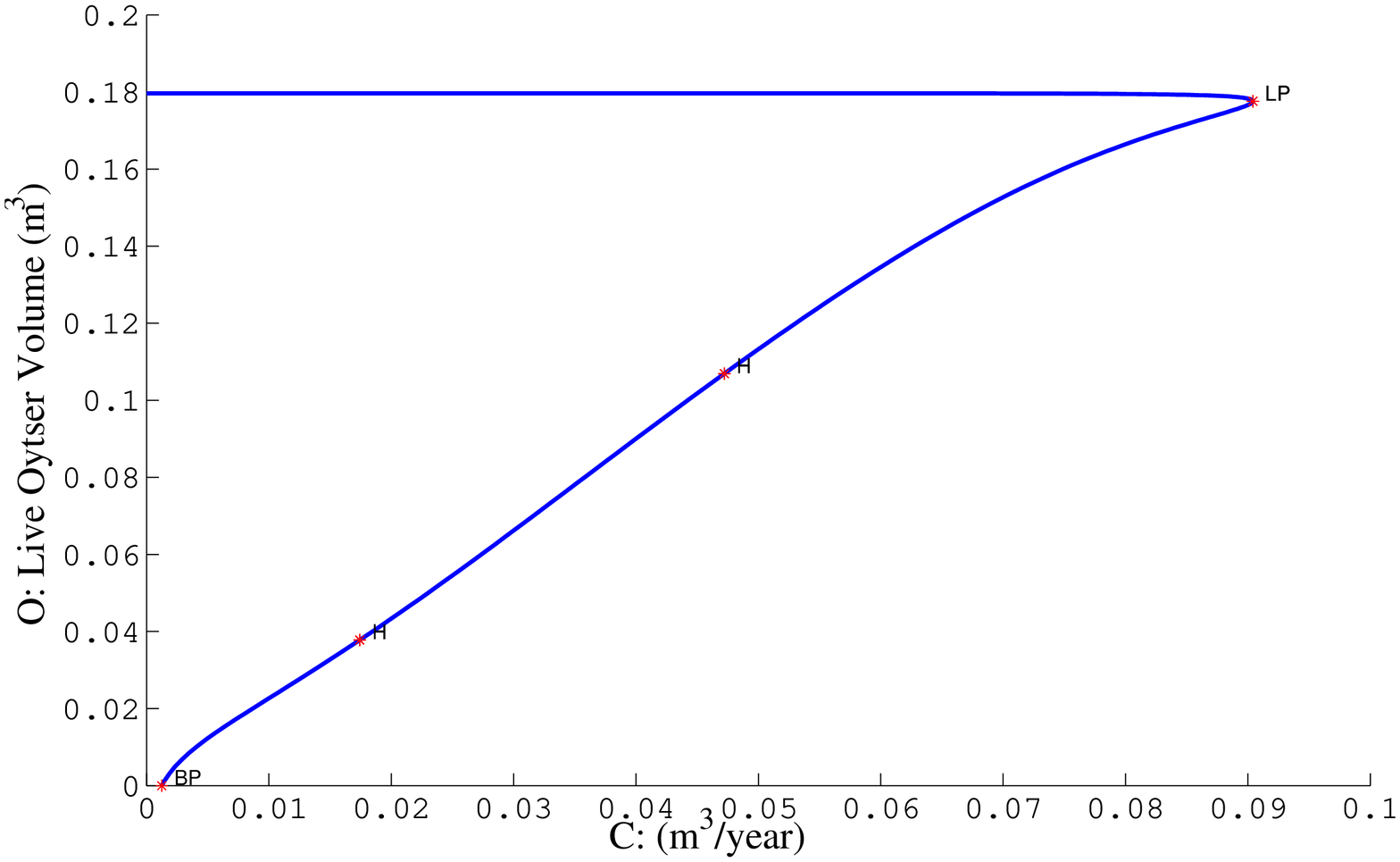}\includegraphics[width=0.5\textwidth]{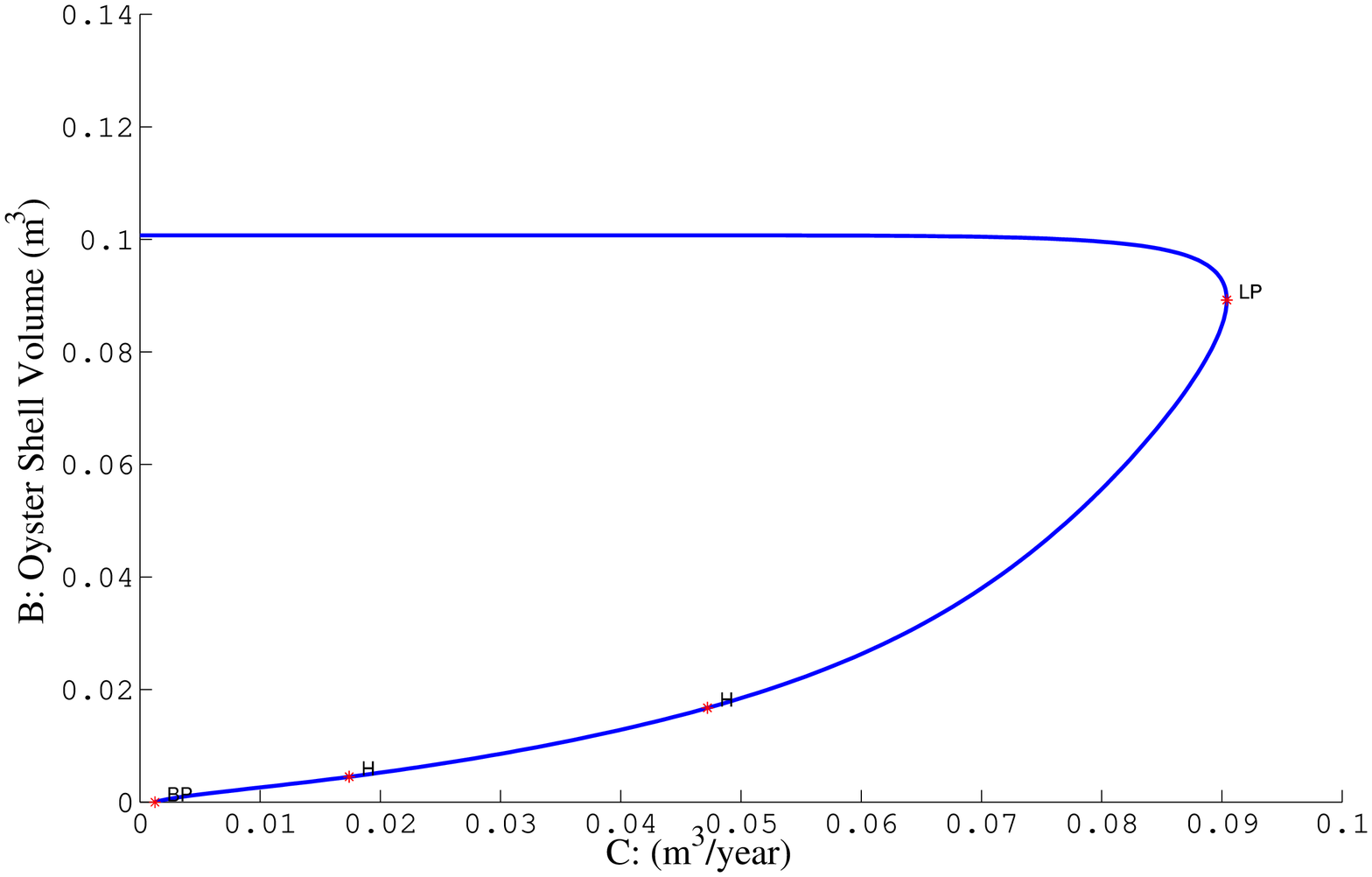}
  \includegraphics[width=0.5\textwidth]{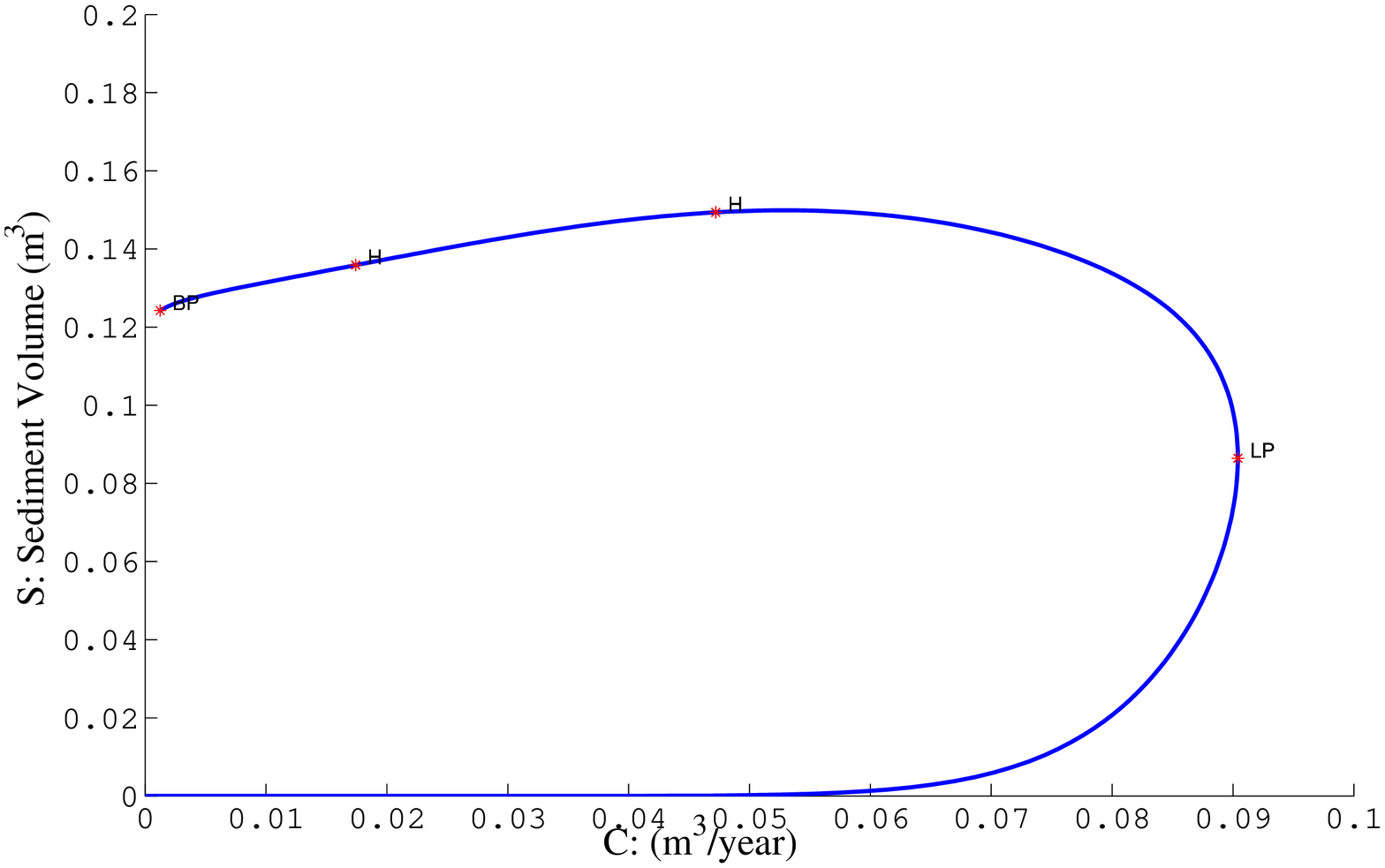}
  \caption{\label{g:b6} Bifurcation diagram of \eqref{7}-\eqref{9}, with parameters given in Table \ref{table1} except $\epsilon=0.05$. Here the horizontal axis is $C$, and the vertical axis is (upper left) $O$, (upper right) $B$ and (lower) $S$.}
\end{figure}

The system can shift from Case 2 to Case 3, where a transcritical bifurcation occurs for a positive $C_*(r)$ (Fig.~\ref{g:b6}) when $r$ is large or when $\epsilon$ is low.

\section{Conclusions}\label{sec6}

\subsection{Key findings}

We constructed an ordinary differential equation model for the volumes of live oysters, shell from dead oysters, and accumulated sediment on an oyster reef. Feedback interactions between the oysters and the sediment occurred such that low-relief (\textit{i.e.}, low vertical height) reefs were eventually choked by sediment, whereas high-relief reefs had less sediment deposition due to their height off the sea floor and filtration of the sediment by live oysters. Using sediment deposition rate as a bifurcation parameter, we have shown that the oyster-free (\textit{i.e.}, degraded reef) equilibrium point is stable for sufficiently high sediment influx. In contrast, it is unstable at lower sediment deposition rates, and in that case the system has at least one equilibrium point with live oysters and a persistent reef matrix. Furthermore, the system has a backward bifurcation in sediment concentration when certain criteria are met, whereby there are two stable states, one with live oysters and persistent reef, and one without live oysters or shell reef. Bistability becomes more likely when the oyster reefs grow higher from the sea floor and degrade less quickly, and when the reefs more strongly reduce sediment deposition either through reef height or through enhanced filtering of sediment by live oysters.

We also observed bistability in numerical simulations for physically realistic parameter values. In this case, the long-term fate of a reef depended on its initial conditions. In particular, an initial low volume of dead shells (\textit{i.e.}, reef height) led to extinction of the live oysters and degradation of the reef, whereas an initial high volume of dead shells led to a stable steady state with living oysters. These results are analogous to the experimental field results for low-relief and high-relief reefs, respectively \citep{Schulte09}. This study is therefore the first to provide a theoretical foundation for the existence of bistability in oyster reefs.

\subsection{Model assumptions and caveats}

The assumptions made in our model are the following. Oysters grow logistically and die linearly
when above the sediment. They die at a constant rate per oyster volume when buried by sediment.
To quantify the extent to which oysters are above the sediment, we use a monotonically increasing
function $f\in[0,1]$ that depends on the difference between the reef height and the sediment height. Dead shells in the reef are degraded at a constant rate per dead reef volume. Sediment is eroded at a constant rate per sediment volume. It is deposited at a rate that decreases with the height of the reef in the water column and decreases exponentially with the rate at which oysters filter sediment. The filtration is assumed to be proportional to the oyster volume and to a filtration rate that has a single maximum at some optimal sediment concentration. These assumptions are sufficient to generate regions of bistability. For the numerical studies, we further assumed a particular sigmoidal form for $f$, we assumed that sediment concentration decreases exponentially with height in the water column, and we assumed that the filtration rate is a Ricker-type function of the sediment concentration.

For direct comparison of our results with experimental field data from natural systems (\textit{e.g.}, Chesapeake Bay \citep{Schulte09}), reliable estimates are needed for oyster growth and parameters related to the interaction between oysters and sediment. Further refinements to the model may include seasonal effects, spatial effects such as sources of oyster larvae from distant reefs, and a more accurate representation of reef geometry, including how sediment intercalates between shells.

\subsection{Relevance for oyster reef restoration}

In previous mathematical and conceptual models of oyster reef dynamics, the possibility of alternative stable states was ignored \citep{Powell06, Mann07}, despite the evidence for nonlinear processes in oyster ecology \citep{Cerco07}. This omission promoted a narrow focus regarding the optimal reef architecture in oyster restoration efforts within Chesapeake Bay, despite the highly variable environmental conditions (\textit{e.g.}, sediment deposition) throughout the bay's waters. The collective findings of our theoretical analysis and previous field experiments \citep{Schulte09} demonstrate that the optimal oyster reef architecture will differ based on environmental conditions, and that initial reef height is a critical feature of reef architecture, which determines the eventual persistence or degradation of constructed oyster reefs. These results are not surprising given the existence of alternative stable states in other bivalve mollusks such as the horse mussel  \textit{Atrina zelandica} \citep{Coco06} and blue mussel \citep{Petraitis09}, and in a diverse suite of ecosystems \citep{Scheffer01}, and underscore the need to consider the phenomenon of alternative stable states in ecological restoration. This investigation thus provides a conceptual framework for alternative stable states in native oyster populations, and can be used as a tool to improve the likelihood of success in ecological restoration.

\section*{Acknowledgements}

Partial funding was provided by NSF grants EF-0436318, DMS-0703532 and DMS-1022648, and by a grant from the US Army Corps of Engineers, Norfolk District. RNL thanks D. Schulte, R. Burke, S. Jordan and R. Seitz for insightful comments on oyster ecology. The authors thank the editor and three anonymous reviewers for various suggestions that improved the manuscript. This manuscript resulted from an Honor's Thesis completed at the College of William \& Mary by WJC.

%
%
%
%

\section*{Appendix: Calculation of the linearization and stability}

In this section we explain the method of bifurcation of equilibrium points from a known branch of trivial equilibria. It is well-known as \lq\lq bifurcation from a simple eigenvalue\rq\rq\ in the studies of analytical bifurcation theory (see \citep{Crandall71, Jiang09, Liu07}). Here we apply this powerful method to a finite-dimensional problem.

Consider a smooth mapping $F=F(\la,u):\R\times U\to \R^n$ where $U$ is an open subset of $\R^n$, $n\ge 1$, $\la$ is a parameter and $u$ is the state variable. We consider the equilibrium problem
\begin{equation}\label{5.1}
    F(\la,u)=0.
\end{equation}
Assume that a trivial solution is known. That is, there exists $u_0\in U$ so that $F(\la,u_0)=0$ for all $\la\in \R$. So $\{(\la,u_0):\la\in \R\}$ is a line of trivial solutions of \eqref{5.1}.

The linearization of $F$ with respect to $u$ is represented by the Jacobian matrix: $F_u=(J_{ij}=\partial_j F_i)$, where $J_{ij}$ is the entry of $F_u$ at row $i$ and column $j$, and
\begin{equation*}
\partial_j F_i=\frac{\partial F_i(\la,u)}{\partial u_j}, \;\;\; 1\le
i,j\le n,
\end{equation*}
is the partial derivative. Note that $F=(F_1, F_2, \cdots, F_n)$ and $u=(u_1,u_2,\cdots,u_n)$ are both vectors in $\R^n$. Similarly the second derivative of $F$ on $u$ is expressed as a $3$-dimensional matroid $F_{uu}=(K_{ijk}=\partial_{jk} F_i)$, where
\begin{equation*}
\partial_{jk} F_i=\frac{\partial^2 F_i(\la,u)}{\partial u_j\partial u_k}, \;\;\; 1\le
i,j,k\le n,
\end{equation*}
is the second order partial derivative. Also the mixed derivative $F_{\la u}=(M_{ij}=\partial_{\la j} F_i)$ where
\begin{equation*}
\partial_{\la j} F_i=\frac{\partial^2 F_i(\la,u)}{\partial u_j\partial \la}, \;\;\; 1\le
i,j\le n.
\end{equation*}
We notice that $F_u$ defines a linear operator $\R^n\to\R^n$ with matrix multiplication and so does $F_{\la u}$.  $F_{uu}$ defines a bilinear operator $\R^n\times \R^n\to \R^n$ which can be expressed as
\begin{equation*}
    F_{uu}[(x_1, \cdots, x_n),(y_1, \cdots,
    y_n)]=(\sum_{j,k}K_{1jk}x_jy_k, \cdots,
    \sum_{j,k}K_{njk}x_jy_k).
\end{equation*}
Finally for a linear operator $L:\R^n\to\R^n$, we use $N(L)$ and $R(L)$ to denote the null space and the range of $L$; and we use $\langle x,y\rangle$ to denote the standard dot product of $x,y\in\R^n$.

Now we are ready to state a bifurcation theorem due to Crandall and Rabinowitz \citep{Crandall71} (here we only state a special case):
\begin{theorem}
Let $F:\R\times U\to \R^n$ be twice continuously differentiable, where $U$ is an open subset of $\R^n$. Suppose that $F(\la, u_0)=0$ for $\la \in \R$, and at $(\lambda_0, u_0)$, $F$ satisfies
\begin{enumerate}
\item[{\bf(F1)}]
$dim N(F_u(\la_0, u_0))=codim R(F_u(\la_0,u_0))=1$, and
$N(F_u(\la_0, u_0))=Span\{w_0\}$;
\item[{\bf (F3)}] $F_{\la
u}(\la_0,u_0)[w_0] \not\in R(F_u(\la_0,u_0)).$
\end{enumerate}
\noindent  Then  the solutions of \eqref{5.1} near $(\lambda_0, u_0)$ consists precisely of the curves $u=u_0$ and $(\la(s),u(s))$, $s \in I=(-\delta,  \delta)$, where $(\la(s),u(s))$ are continuously differentiable functions such that $\la(0)=\la_0$, $u(0)=u_0$, $u'(0)=w_0$. Moreover
\begin{equation} \label{2.13} \la'(0)=-\frac{\langle l,
F_{uu}(\la_0,u_0)[w_0,w_0]\rangle}{2\langle l, F_{\la
u}(\la_0,u_0)[w_0]\rangle},\end{equation}
where $l\in \R^n$
satisfying $R(F_{u}(\la_0,u_0))=\{y\in \R^n: \langle
l,y\rangle=0\}$.
\end{theorem}
In simpler terms, at a bifurcation point $\la=\la_0$, the Jabobian $F_u$ has zero as an eigenvalue; {\bf(F1)} means that zero is a simple eigenvalue of $F_u$, which means that the eigenspace of $F_u$ is one-dimensional, and the range of $F_u$ is $(n-1)$-dimensional (called, codimension one); {\bf(F3)} means that $F_{\la u}[w_0]$ does not belong to the range of $F_u$, where $w_0$ is any nonzero eigenvector. Once these conditions are satisfied, then there is a curve of solutions bifurcating from the branch of trivial solutions. The formula of $\la'(0)$ is useful for determining the direction of the bifurcation (forward/backward).

To apply the above abstract theory to our problem, we notice that the trivial equilibrium $(C,O,B,S)=(C,0,0,C/\beta)$ is not constant for $C$. Hence we make a change of variable $z=S-C/\beta$ then $(O,B,z)=(0,0,0)$ is a constant solution. We define
\begin{equation}\label{5.2}
    G(C,O,B,z)=\left(
                 \begin{array}{c}
                   \ds r O f(d) \left(1 - \frac{O}{k}\right)-\mu f(d) O - \epsilon(1 - f(d)) O \\
                   \ds r f(d)\frac{O^2}{k} + \mu f(d) O - \gamma B + \epsilon(1 - f(d)) O \\
                   \ds Cge^{-\frac{FO}{Cg}} - \beta z-C\\
                 \end{array}
               \right),
\end{equation}
where $d=\la (O/2+B)-z-C/\beta$, and definitions of $f,k,g,F$ are same as before. Let $u=(O,B,z)$. Then $G_u(C,0,0,0)$ is the same as \eqref{20}. At $C=C_*$, $G_u(C_*,0,0,0)$ can be written as
\begin{equation}\label{5.3}
   L\equiv G_u(C_*,0,0,0)= \left(
  \begin{array}{ccc}
    0 & 0 & 0 \\
    \ds\frac{\epsilon r}{r-\mu+\epsilon} & -\gamma & 0 \\
   C_* g'(0) - F(C_*) & C_* g'(0)  & -\beta \\
  \end{array}
\right).
\end{equation}
We take the eigenvector of $L$ to be $w_0=(1, w_{02}, w_{03})$ where
\begin{equation*}\begin{split}
    &w_{02}=\frac{\epsilon r}{\gamma(r-\mu+\epsilon)}, \\
    &w_{03}=\frac{C_* g'(0) - F(C_*)}{\beta}+\frac{C_* g'(0)\epsilon r}{\beta \gamma(r-\mu+\epsilon)},
\end{split}\end{equation*}
one can see that the range of $L$ is $\{(0,y,z)\in \R^3\}$ which is two-dimensional, so we can take the vector $l$ to be $(1,0,0)$. A vector $v$ does not belong to the range of $L$ if the first entry of $v$ is not zero. So to apply \eqref{2.13}, we \textit{only} need to calculate the derivatives from the first equation of the system.

We can calculate that $\langle l, G_{\la u}(C_*,0,0,0)[w_0]\rangle = -f'(-C_*/\beta)(r-\mu+\epsilon)/\beta<0$, and with a more tedious
calculation, we find that
\begin{equation*}
\begin{split}
\langle l,
&G_{uu}(_*,0,0,0)[w_0,w_0]\rangle\\
=&\frac{2r}{k}f'(-C_*/\beta)\left[
\frac{\la
(r-\mu+\epsilon)k}{2r}-\frac{f(-C_*/\beta)}{f'(-C_*/\beta)}
+\ds\frac{\la \epsilon k}{\gamma}-\frac{\epsilon k}{\gamma \beta}\left(C_* g'(0) - F(C_*)+\frac{C_* g'(0)\epsilon r}{\gamma(r-\mu+\epsilon)}\right)\right].
\end{split}
\end{equation*}

Hence combining all the calculations, we obtain the direction of the bifurcating curve as
\begin{equation*}
    C'(0)=-\frac{\langle l,G_{uu}(C_*,0,0,0)[w_0,w_0]\rangle}{2\langle l, G_{\la u}(C_*,0,0,0)[w_0]\rangle}=\frac{rI}{k(r-\mu+\epsilon)}.
\end{equation*}

\noindent\textbf{\Large References}


\bibliographystyle{elsarticle-harv}
\bibliography{RefsJTheorBiol1a}
\end{document}